\documentclass[10pt,prb,superscriptaddress,amsfonts,amssymb,floats,amsmath,twocolumn,aps,footinbib,shownopacs,longbibliography]{revtex4-2}

\usepackage{amsmath}
\usepackage{amsfonts}
\usepackage{graphicx}
\usepackage{color}
\usepackage{xcolor}

\usepackage{hyperref}
\usepackage{soul}
\usepackage{braket}
\usepackage{cleveref}

\usepackage{cancel}

\usepackage[normalem]{ulem}   

\usepackage{tikz}

\begin{document}
\title{Distinguishing Majorana bound states from accidental zero-energy modes with a microwave cavity}
\author{Sarath Prem}
 \email{sarathprem@MagTop.ifpan.edu.pl}
\affiliation{International  Research  Centre  MagTop,  Institute  of  Physics,  Polish  Academy  of  Sciences, Aleja  Lotnikow  32/46,  PL-02668  Warsaw,  Poland}
\author{Olesia Dmytruk}
\email{olesia.dmytruk@polytechnique.edu}
\affiliation{CPHT, CNRS, École polytechnique, Institut Polytechnique de Paris, 91120 Palaiseau, France}
\author{Mircea Trif}
 \email{mtrif@MagTop.ifpan.edu.pl}
\affiliation{International  Research  Centre  MagTop,  Institute  of  Physics,  Polish  Academy  of  Sciences, Aleja  Lotnikow  32/46,  PL-02668  Warsaw,  Poland}
\affiliation{Department of Physics, University of Basel, Klingelbergstrasse 82, 4056 Basel, Switzerland}

\date{\today}

\begin{abstract} 
Transport measurements of hybrid nanowires often rely on the observation of a zero-bias conductance peak as a hallmark of Majorana bound states (MBSs). However, such signatures can also be produced by trivial zero-energy Andreev bound states (ABSs) or by quasi-Majorana bound states (QMBSs), complicating their unambiguous identification. Here we propose microwave absorption visibility, extracted from parity-dependent cavity-nanowire susceptibility measurements, as a complementary probe of MBSs nonlocality. We study a Rashba spin-orbit nanowire consisting of a proximitized superconducting segment and an uncovered quantum-dot region, capacitively coupled to a single-mode microwave cavity. We show that true MBSs yield finite visibility only when both MBSs are simultaneously coupled to the cavity, reflecting their intrinsic nonlocality. In contrast, ABSs and QMBSs exhibit visibility extrema even when the cavity couples only locally to part of the nanowire. We further demonstrate that this distinction persists in the presence of Gaussian disorder, which may otherwise generate trivial subgap states. Motivated by recent experiments, we also analyze ``poor man's'' Majoranas in double-quantum-dot setups, where analytical results confirm the same nonlocal visibility criterion. Finally, we discuss a cavity-driven scheme for  initializing the electronic system in a given parity state. Our results establish cavity-based visibility as a robust and versatile probe of MBSs, providing a clear route to distinguish them from trivial zero-energy states in hybrid superconducting platforms. 
\end{abstract}

\maketitle

\section{Introduction}
In condensed-matter physics, Majorana bound states (MBSs) are localized zero-energy quasiparticles at the boundaries of topological superconductors, where spatially separated pairs of MBSs form a nonlocal fermionic degree of freedom 
~\cite{Alicea_Rep.Prog.Phys..2012New,Beenakker_Annu.Rev.Condens.MatterPhys..2013Search,Lutchyn_NatRevMater.2018Majorana,Prada_NatRevPhys.2020Andreev,Laubscher_J.Appl.Phys..2021Majorana}. Their non-Abelian statistics make them promising candidates for fault-tolerant topological quantum computation \cite{Nayak_Rev.Mod.Phys..2008NonAbelian,Stern_Science.2013Topological}. MBSs are predicted to emerge in various condensed-matter platforms, including topological insulators in proximity to an $s$-wave superconductor
~\cite{Fu_Phys.Rev.Lett..2008Superconducting,Fu_Phys.Rev.B.2009Josephson, Cook_Phys.Rev.B.2011Majorana,
Schulz_Phys.Rev.Res..2020Majorana, Legg_Phys.Rev.B.2021Majorana}, 
superconductor-semiconductor heterostructures~\cite{Sau_Phys.Rev.Lett..2010Generic, Alicea_Phys.Rev.B.2010Majorana}, graphene-based platforms
~\cite{Klinovaja_Phys.Rev.Lett..2012Electricfieldinduced,
Klinovaja_Phys.Rev.B.2012Helical,
Black-Schaffer_Phys.Rev.Lett..2012Edge,
Klinovaja_Phys.Rev.X.2013Giant,
San-Jose_Phys.Rev.X.2015Majorana,
Dutreix_Eur.Phys.J.B.2014Majorana,
Kaladzhyan_SciPostPhys..2017Formation,Marganska_Phys.Rev.B.2018Majorana}, and chains of magnetic adatoms on a superconductor 
\cite{Nadj-Perge_Phys.Rev.B.2013Proposal,Klinovaja_Phys.Rev.Lett..2013Topological,Braunecker_Phys.Rev.Lett..2013Interplay,Vazifeh_Phys.Rev.Lett..2013SelfOrganized,
Peng_Phys.Rev.Lett..2015Strong}.

Among the various platforms proposed for realizing MBSs, one-dimensional (1D) semiconducting nanowires with strong Rashba spin-orbit coupling, proximity-induced superconductivity, and subject to an external magnetic field ~\cite{Lutchyn_Phys.Rev.Lett..2010Majorana,Oreg_Phys.Rev.Lett..2010Helical} have garnered significant theoretical and experimental attention. Zero-bias peaks in the local conductance measurements are one of the signatures of the MBSs
~\cite{Law_Phys.Rev.Lett..2009Majorana,Flensberg_Phys.Rev.B.2010Tunneling,Lin_Phys.Rev.B.2012Zerobias,DasSarma_Phys.Rev.B.2012Splittinga}that have been  observed in multiple experiments
~\cite{Mourik_Science.2012Signatures,Deng_NanoLett..2012Anomalous,Churchill_Phys.Rev.B.2013Superconductornanowire,
Deng_Science.2016Majorana,
Das_NaturePhys.2012Zerobias, 
DeMoor_NewJ.Phys..2018Electric}. However, it was theoretically shown that 
they could also be a result of disorder~\cite{Liu_Phys.Rev.Lett..2012ZeroBias} or 
the presence of zero-energy Andreev bound states (ABSs) in setups with a normal part~\cite{Liu_Phys.Rev.B.2017Andreev,Reeg_Phys.Rev.B.2018Zeroenergy,Hess_Phys.Rev.B.2021Local,Prada_NatRevPhys.2020Andreev,Hess_Phys.Rev.Lett..2023Trivial,Sahu_Phys.Rev.B.2023Effect}. Furthermore, nonlocal differential conductance allows one to probe the closing and reopening of the bulk gap, which is usually associated with the topological phase transition and was measured
~\cite{Aghaee_Phys.Rev.B.2023InAsAl}. But it was theoretically shown that multiple ABSs in the bulk could also produce trivial gap reopening signatures
~\cite{Hess_Phys.Rev.Lett..2023Trivial}.

Another source of zero-bias peaks in the local conductance is the smooth  inhomogeneities  in the semiconductor-superconductor platforms~\cite{Kells_Phys.Rev.B.2012Nearzeroenergy,Prada_Phys.Rev.B.2012Transport,Penaranda_Phys.Rev.B.2018Quantifying}. Such inhomogeneities give rise to the formation of local quasi-MBSs (QMBSs)~\cite{Vuik_SciPostPhys..2019Reproducing}. Despite lacking true nonlocality—a defining feature of true MBSs—QMBSs can nonetheless appear as zero-energy excitations in spectroscopic measurements. This underscores their role as a significant source of zero-bias conductance peaks, complicating the experimental differentiation between topological and trivial origins~\cite{Prada_NatRevPhys.2020Andreev}.

This motivates the development of alternative approaches to access other properties of the MBSs. Cavity quantum electrodynamics (cQED) platforms—where the hybrid nanowire is capacitively coupled to a high-quality microwave cavity—have emerged as a promising route. Such systems have been shown to detect the topological phase transition point \cite{Trif_Phys.Rev.Lett..2012Resonantly,Dmytruk_Phys.Rev.B.2015Cavity,Dmytruk_Phys.Rev.B.2016Josephson}, reveal the self-adjoint nature of MBSs, and probe fermion parity \cite{Cottet_Phys.Rev.B.2013Squeezing,Dartiailh_Phys.Rev.Lett..2017Direct,Cottet_J.Phys.:Condens.Matter.2017Cavity,Trif_Phys.Rev.B.2018Dynamic,Trif_Phys.Rev.Lett..2019Braiding,Dmytruk_Phys.Rev.B.2024Hybrid} (also see \cite{Ricco_SciRep.2022Accessing,Ricco_Phys.Rev.A.2022Reshaping}). Recently, fermion parity readout was demonstrated experimentally through quantum capacitance measurements \cite{MicrosoftAzureQuantum_Nature.2025Interferometric}; however, this scheme required coupling an additional elongated quantum dot to the topological wire.
By contrast, cQED schemes employing microwave cavities can directly probe the hybrid wire itself, without relying on ancillary degrees of freedom. In particular, they provide access to the nonlocal characteristics of MBSs by exploiting their coupling to extended (gapped) states of the wire \cite{Dmytruk_Phys.Rev.B.2023Microwave}.
 
\begin{figure}[t]
    \centering
    \includegraphics[
    trim=0cm 4cm 0cm 4cm, 
    width=\linewidth]{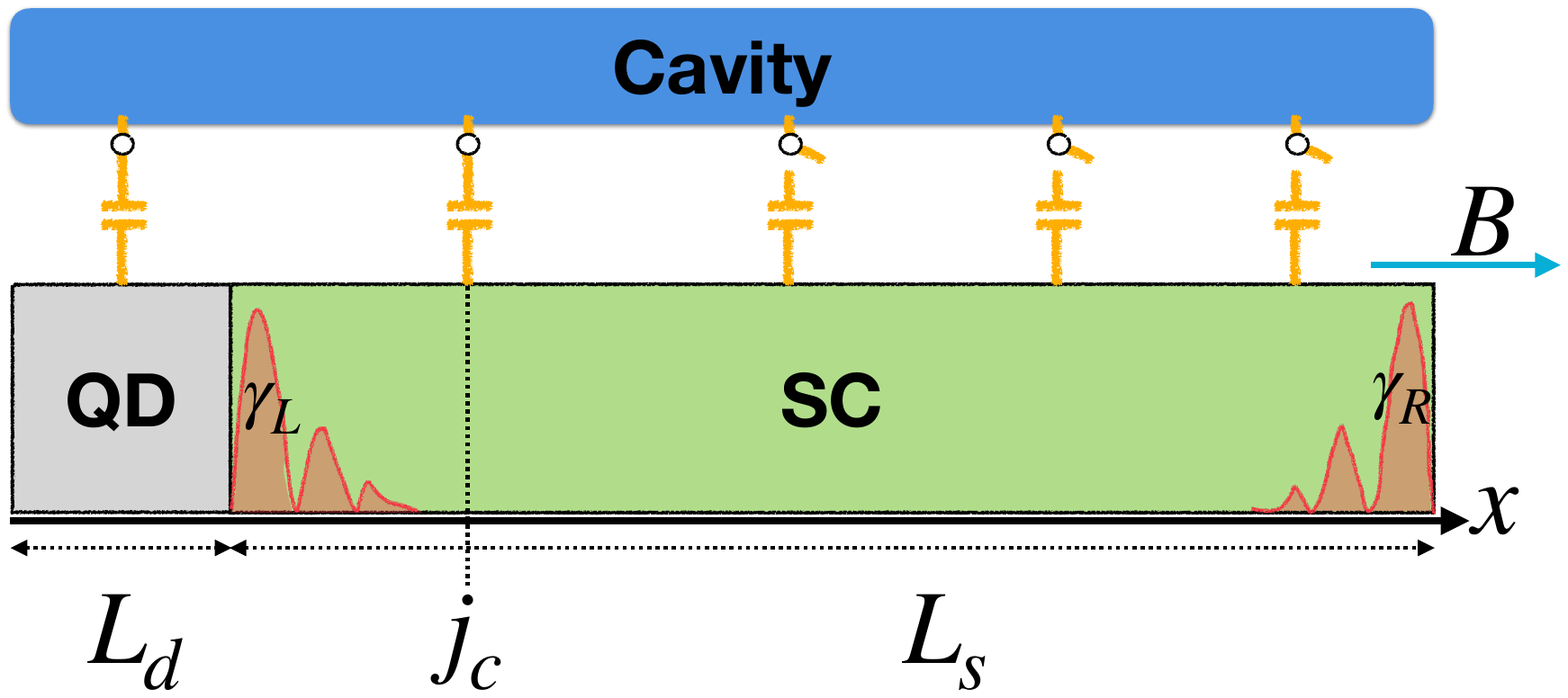}
    \caption{
Schematic of a Rashba nanowire (gray) consisting of \( L = L_d + L_s \) lattice sites, partially covered by an \( s \)-wave superconductor (SC, green) and subjected to a magnetic field \( B \) oriented along the positive \( x \) axis. The uncovered segment comprising the first \( L_d \) sites defines the quantum dot (QD) or normal region. The remaining \( L_s \) sites (green) form the proximized superconducting region, where pairing is induced by an adjacent \( s \)-wave SC (not shown). The nanowire is capacitively coupled (yellow) to a one-dimensional microwave cavity (blue), with the coupling extending up to a specific site \( j_c \) along the wire. By tuning system parameters—such as the chemical potential and the Zeeman splitting induced by the magnetic field $B$—the system can be driven into a topological SC phase, where zero-energy MBSs $\gamma_{L,R}$ appear localized at the ends of the topological SC (shown in red). The presence of these electronic excitations alters the cavity’s frequency $\omega_c$ and decay rate $\kappa$, which can be probed through dispersive readout methods.
}
    \label{SystemSchematic}
\end{figure}

In this work, we propose to use the cQED platform to probe the nonlocality of MBSs and to differentiate MBSs from other zero-energy states, such as ABSs and QMBSs. Specifically, 
we study a Rashba nanowire with an axial magnetic field that is partially covered by an $s$-wave superconductor and 
capacitively coupled to a microwave cavity (see Fig. \ref{SystemSchematic}). 
To probe the localization of the bound states within the wire, we consider selective site-dependent coupling of the cavity with the wire that can be implemented as an extension of the experiment in Ref.~\cite{Bruhat_Phys.Rev.X.2016Cavity}.
Microwave measurements provide access to the electronic susceptibility of the wire, with its imaginary and real parts corresponding to the absorption and cavity frequency shift, respectively ~\cite{Dmytruk_Phys.Rev.B.2015Cavity,
Dartiailh_Phys.Rev.Lett..2017Direct,
Cottet_J.Phys.:Condens.Matter.2017Cavity,
Dmytruk_Phys.Rev.B.2023Microwave}. For a 1D  topological superconductor, it was found that the electronic susceptibility—both the real and imaginary parts—is different for even and odd Majorana parities only if both MBSs localized at the opposite ends of the superconductor are coupled to a cavity~\cite{Dmytruk_Phys.Rev.B.2023Microwave}. The extent of nonlocality in bound states can be more precisely established by employing the visibility of the microwave absorption, which is simply referred to as visibility throughout this work. This approach was previously applied in an investigation of Majorana modes forming in chains of magnetic adatoms placed on superconductors and interacting with the surrounding spin waves (or magnons) \cite{Shen_Phys.Rev.Research.2023Majoranamagnon}.

This work extends upon the foundational framework established by prior studies, unveiling three key novel contributions: (i) we evaluate the microwave absorption visibility pertaining to the parity of the zero-energy bound states existing within the system; (ii) we provide strong evidence that this measure effectively distinguishes between MBSs, zero-energy ABSs, and QMBSs; and (iii) we utilize our framework to explore the emergence of ``poor man's'' Majoranas (PMMs) in minimalistic settings involving superconducting QDs.
Although the first two points primarily require numerical computations (including the effect of disorder), the last point allows us to formulate straightforward and insightful analytical expressions that can be beneficial for future experiments aimed at revealing PMMs. The properties of various PMMs can be verified or disproven by integrating them with current transport measurement techniques. Beyond these core results, our approach naturally generalizes to more complex architectures: By considering multiple Majoranas along a single nanowire or in networks of coupled nanowires, cavity visibility could serve as a diagnostic of nonlocal correlations among several MBSs and as a probe of collective topological modes. These extensions suggest that cavity-based visibility can provide not only a clear identification of MBSs but also a route toward their manipulation in larger hybrid superconducting networks.

The structure of the paper is as follows: In Sec.~\ref{modelHamiltonian} we introduce the model Hamiltonian and compile the important quantities for this work. Section~\ref{mainresults} elaborates on the principal findings related to how the detection of microwave absorption can serve as a means to differentiate MBSs from ABSs and QMBSs. Subsequently, in Sec.~\ref{EffectOfDisorder}, we provide an in-depth examination of how disorder influences the detectability of microwave absorption. Furthermore, Sec.~\ref{barriersection} offers a detailed analysis of the impact that potential barriers have on the visibility of this absorption. Next, in Sec.~\ref{poormanMajorana}, we apply the idea of visibility to implementations supporting the PMMs and discuss a scheme that allows the initialization of parity in the presence of an external drive. We summarize our findings and provide an outlook for future work in Sec.~\ref{conclusion}.

\begin{table*}[t]
\begin{ruledtabular}
\begin{tabular}{cllcccc}
   Regime 
   &   $\mu_{d}/t_{s}$ &$\alpha_{d}/t_{s}$&$\mu_{s}/t_{s}$&  $\alpha_{s}/t_{s}$ &  $\Delta_{\text{min}}/t_{s}$&  $\epsilon_{0}/\Delta_{\text{min}}$\\
   \hline
   Pristine MBS ($L_d=0$)
   &   --- & --- &0&  0.555 &  0.012&  $5.12\times 10^{-5}$\\[6pt]
   MBS  
   &   0 & 1.439&0&  0.555 &  0.012&  $3.25\times 10^{-5}$\\[6pt]
   ABS  
   &   0 &1.439&  0.1 &  0 &  0.0125&  $2.76\times10^{-4}$\\[6pt]
   QMBS 
   &   0 & 1.439 &0&  0.555 &  0.012&  $9.52\times10^{-6}$\\[6pt]
\end{tabular}
\end{ruledtabular}
\caption{We consider $L_{d}=60$ sites for the QD  and $L_{s}=500$ sites  for the SC, giving $L=560$ as total sites (pristine MBS has no QD). Here
 $\mu_{d}$ ($\mu_{s}$) is the chemical potential in the QD (SC),
 $t_{d}/t_s=5\,(t_{s}=20\, \text{meV})$ is the hopping amplitude in the QD (SC), $\alpha_{d}$ ($\alpha_{s}$) is the Rashba spin-orbit strength in the QD (SC), 
 $E_{Z,d}/t_{s}=0.06875 $ ($=E_{Z,s}/t_{s}$) is the Zeeman energy in the QD (SC), $\Delta/t_{s}=0.0125$ is the proximity‐induced superconducting gap in the SC, $\Delta_{\rm min}$ is the smallest bulk gap in the periodic‐boundary SC, and $\epsilon_{0}$ is zero‐mode energy. For all plots, we chose a linewidth of $\eta/\Delta = 4 \times 10^{-3}$, which is smaller than the average level spacing.
}
\label{parametertable}
\end{table*}

\section{Model Hamiltonian}\label{modelHamiltonian}

We consider a  Rashba nanowire in an external magnetic field parallel to the wire, as depicted in Fig. \ref{SystemSchematic}.  The first $L_{d}$ sites from the left of the wire define a quantum dot (QD in Fig. \ref{SystemSchematic}), also referred to as the normal part. The remaining $L_s$ sites of the wire exhibit proximity-induced superconductivity caused by a nearby $s$-wave superconductor (SC). The wire is coupled to a microwave cavity resonator via capacitors (shown in yellow) that can be turned on and off to achieve site-dependent cavity-wire coupling. The total Hamiltonian \cite{Reeg_Phys.Rev.B.2018Zeroenergy,Dmytruk_Phys.Rev.B.2023Microwave} takes the form $H_{\rm tot} = H_{\rm el} + H_{c} + H_{\rm el-c}$, where
\begin{align}
    &H_{\rm el}= \quad \sum_{j=1}^{L}\sum_{\sigma, \sigma'=\uparrow,\downarrow}  c_{j,\sigma}^\dagger
    \left[ 
    \left(
    2 t_{j} - \mu_j
    \right) \delta_{\sigma \sigma'} 
    + E_{Z,j} \, \sigma^x_{\sigma \sigma'}
    \right] c_{j,\sigma'} \nonumber
    \\
    & \quad- \sum_{j=1}^{L-1}\sum_{\sigma, \sigma'=\uparrow,\downarrow }  \left[
    c_{j+1, \sigma}^\dagger 
   \left( t_{j} \delta_{\sigma \sigma'}- i \alpha_j \sigma^y_{\sigma \sigma'}\right)
    c_{j, \sigma'}
    + {\rm H.c.}
    \right]\nonumber\\
    & \quad + \sum_{j=1}^{L} \left( 
    \Delta_j \, c_{j,\uparrow}^\dagger c_{j,\downarrow}^\dagger 
    + {\rm H.c.}
    \right),  
    \label{electronicparttightbindingHamiltonian}
\end{align}
is the (tight-binding) Hamiltonian of the electronic part only for a total wire length $L$, $ H_c = \omega_c a^\dagger a$ represents the photons in the cavity, assumed to be a single mode with a frequency $\omega_c$, and 
\begin{align}
     H_{\rm el-c} &= \sum_{j=1}^{L} \sum_{\sigma=\uparrow,\downarrow}\, g_j c_{j,\sigma}^\dagger  c_{j,\sigma}\,(a^\dagger + a)
     \label{electroncavityinteraction}
\end{align}
describes the capacitive coupling between electrons within the wire and the cavity, as detailed in Ref.~\cite{Dmytruk_Phys.Rev.B.2023Microwave}. Here $g_j = g\,\theta(j)\theta(j_c-j)$ denotes the spatially dependent coupling strength, where $j_c$ represents the final site interacting with the cavity, $g$ is the strength of the coupling, and $\theta(x)$ is the Heaviside function. In addition, $a^\dagger$ ($a$) denotes the creation (annihilation) operator for the photons within the cavity, while  $c_{j,\sigma}^\dagger$ ($c_{j,\sigma}$) is the creation (annihilation) operator for an electron with spin $\sigma=\uparrow, \downarrow$ at position $j$. Moreover,  $\mu_j$ is the  spatially dependent chemical potential, while $E_{Z,j} = (1/2) g_{B,j} \mu_B B$  is the Zeeman energy pertaining to the axial magnetic field $B$ (refer to Fig. \ref{SystemSchematic}), where $g_{B,j}$ represents the spatially dependent $g$ factor and $\mu_B$ is the Bohr magneton.  The parameter $\alpha_j$ symbolizes the spin-orbit interaction (SOI) for the bond $j,j+1$ in the lattice. The Pauli matrices, $\sigma^{x,y,z}$, operate within the spin space, and $\Delta_j$ is the superconducting gap introduced via proximity at site $j$.  
We consider distinct parameter values for the two regions under investigation. Specifically, $t_j=t_{d}\,(t_s)$ represents the tunneling parameter within the QD (superconducting) region. Similarly, $\alpha_j=\alpha_{d}\,(\alpha_s)$ signifies the SOI coupling strength for the dot (superconducting) region. Additionally, $E_{Z,j}=E_{Z,d}\,(E_{Z,s})$ signifies the Zeeman energy in the dot (superconducting) region, while  $\Delta_j=0$ $(\Delta)$ corresponds to the proximity-induced superconducting gap in the dot (superconducting) region. We further assume that the tunneling and SOI coupling strengths at the interface between the two sites are the average values of those found in the respective regions~\cite{Reeg_Phys.Rev.B.2018Zeroenergy}.  
To distinguish between trivial and topological phases of the superconducting segment, it is useful to define the local quantity $\Delta_{t,j} = E_{Z,j} - \sqrt{\mu_j^2 + \Delta_j^2}$. Here, if $\Delta_{t,j}<0$ within the superconducting region, then the system is in the trivial phase. Conversely, if $\Delta_{t,j}>0$, then it signifies the topological phase with in-gap modes emerging at its ends (which become zero-energy modes for sufficiently large systems) \cite{Alicea_Rep.Prog.Phys..2012New,Laubscher_J.Appl.Phys..2021Majorana}. 

Following the Bogoliubov–de Gennes (BdG) formalism to diagonalize the Hamiltonian $H_{\rm el}$, we write the fermionic field  operator as \cite{Dmytruk_Phys.Rev.B.2023Microwave} 
\begin{equation}
    c_{j,\sigma}^\dagger =  
    \sum_{n= 0 }^{2L -1}\left[u_{n\sigma}^*(j) b_n^\dagger+ v_{n\sigma}(j) b_n\right],
\end{equation}
where $u_{n\sigma}(j)$ [$v_{n\sigma}(j)$] are the electron (hole) coherence factors for the state $n$ at position $j$ with spin $\sigma=\uparrow,\downarrow$. The operators $b^\dagger_n$ ($b_n$) are the creation (annihilation) operators for the Bogoliubons of energy $\epsilon_n$ that diagonalize the electronic Hamiltonian as $H_{\rm el}= \sum_{\epsilon_n \geq 0} \epsilon_n (b_n^\dagger b_n-1/2)$.  By adjusting the variables as shown in Table~\ref{parametertable} and closely examining the value of $\Delta_{t,j}$ in the superconducting region, the existence of zero-energy ($\epsilon_0=0$) states—MBSs, ABSs, or QMBSs—can be revealed within the electronic system. The Bogoliubon describing the zero-mode can be represented by the creation (annihilation) operator $b_0^\dagger$ ($b_0$) so that the two-dimensional degenerate many-body state is described by $|0\rangle$ (vacuum, or $n_0\equiv\langle b_0^\dagger b_0\rangle=0$) and $b_0^\dagger|0\rangle$ (occupied, or $n_0=1$), respectively.  Furthermore, we define the parity operator $\hat{\mathcal{P}}$ for the Majorana, Andreev, and quasi-Majorana cases, with $\hat{\mathcal{P}} = (-1)^{\hat{N}}$, where $\hat{N}=\sum_{j=1;\sigma=\uparrow,\downarrow}^Lc_{j,\sigma}^\dagger c_{j,\sigma}$  is the total particle operator in the wire. Assuming the absence of parity breaking mechanisms such as quasiparticle poisoning, the parity operator possesses two eigenvalues: $\hat{\mathcal{P}}|0\rangle=+|0\rangle$ for odd ($o$) parity and $\hat{\mathcal {P}}b_0^\dagger|0\rangle=-b_0^\dagger|0\rangle$ for even ($e$) parity.  

\subsection*{Electronic susceptibility and the visibility of microwave absorption}

We now discuss the electronic correlations probed by the microwave cavity. The equation of motion for the cavity field in the presence of coupling to the nanowire reads (for more details, see Refs.~\cite{Dmytruk_Phys.Rev.B.2023Microwave,Shen_Phys.Rev.Research.2023Majoranamagnon}):
\begin{align}
    \dot{a}(t)=-i[\omega_c-\chi(\omega_c)]a(t)-\frac{\kappa}{2}a(t)-\sqrt{\kappa}b_{\rm in}(t)\,,
\end{align}
where $\kappa$ is the decay rate of the cavity, $b_{\rm in}(t)$ is the input signal (coherent drive) that probes the cavity, and $\chi(\omega) = (1/2\pi)\int dt e^{i\omega t}\chi(t)$ with 
\begin{align}
    \chi(t)=-i\theta(t)\langle[O_c(t),O_c(0)]\rangle_{\rm eq}\,,
    \label{susc}
\end{align}
being the susceptibility of the electronic system pertaining to the operator $O_c\equiv\sum_{j,\sigma}g_jc_{j,\sigma}^\dagger c_{j,\sigma}$. Above, all operators are represented in the interaction picture, and the notation $\langle\dots\rangle_{\rm eq}$ denotes taking the average with respect to the (thermal) equilibrium of the electronic system in the absence of the coupling to the cavity. The results of this interaction are apparent: The cavity frequency $\omega_c$ is adjusted to $\omega_c'=\omega_c -{\rm Re}\chi(\omega_c)$, while the decay rate $\kappa$ is altered to $\kappa'=\kappa+2{\rm Im}\chi(\omega_c)$. The explicit expression for the susceptibility can be found, for example, using the Keldysh formalism, which we briefly  describe in Appendix \ref{App_Keldysh}. In a stationary regime, we find (disregarding the so called counter-rotating contributions, which do not lead to dissipation):
\begin{widetext}
    \begin{align}
        \chi(\omega)&=i\sum_{p\sigma,j\sigma'}g_pg_j\sum_{n,m}\frac{[u_{n\sigma}(p)v_{m\sigma}^*(p)-u_{m\sigma}(p)v_{n\sigma}^*(p)][u^*_{m\sigma'}(j)v_{n\sigma'}(j)-u^*_{n\sigma'}(j)v_{m\sigma'}(j)]}{\omega-\epsilon_n-\epsilon_m+i\eta}[1-f(\epsilon_n)-f(\epsilon_m)]\nonumber\\
        &+\frac{[u_{n\sigma}(p)u_{m\sigma}^*(p)-v_{m\sigma}(p)v_{n\sigma}^*(p)][u_{m\sigma'}(j)u^*_{n\sigma'}(p)-v_{n\sigma'}(j)v^*_{m\sigma'}(p)]}{\omega+\epsilon_m-\epsilon_n+i\eta}[f(\epsilon_n)-f(\epsilon_m)]\,,
        \label{susc_general}    
        \end{align}
\end{widetext}
where $f(\epsilon_n)$ is the occupation of the state at energy $\epsilon_n$, and $\eta$ quantifies the linewidth of the levels, which, for simplicity, was  assumed to be uniform. In the regime $\omega<2 \Delta_{\rm min}$, with  $\Delta_{\rm min}$ being the smallest bulk gap in the periodic‐boundary SC, only the transitions between the in-gap and the extended modes contribute to the dissipative response. Therefore, from here on, we will consider only these contributions. Additionally, for the sake of clarity, we assume the presence of a single energy in-gap mode $\epsilon_0<\Delta_{\rm min}$. It is worth noting that our expressions can readily be extended to accommodate more complex scenarios, such as transient regimes with time-dependent distribution functions. Nevertheless, such cases are beyond the scope of this work and will be left for future studies.  We obtain the following:
\begin{align}
   \chi_0(\omega)&\simeq ig^2\sum_{n\neq M}\bigg(\frac{|M_{n0}^o|^2}{\omega-\epsilon_n-\epsilon_0+i\eta}[1-f(\epsilon_n)-f(\epsilon_0)]\nonumber\\
        &+\frac{|M_{n0}^e|^2}{\omega+\epsilon_0-\epsilon_n+i\eta}[f(\epsilon_0)-f(\epsilon_m)]\bigg)\,,
\end{align}
where
\begin{align}
\mathcal{M}^e_{n0}(j_c)&\!=\!\!\sum_{p=1}^{j_c}\sum_{\sigma=\uparrow,\downarrow}
\left[
u_{0\sigma}^*(p) u_{n\sigma}(p) -v_{0\sigma}^*(p) v_{n\sigma}(p)
\right]\nonumber\,,\\
\mathcal{M}^o_{n0}(j_c) &\!=\!\!\sum_{p=1}^{j_c}\sum_{\sigma=\uparrow,\downarrow} 
\left[
v_{0\sigma}(p) u_{n\sigma}(p)-u_{0\sigma}(p) v_{n\sigma}(p)
\right]\,,
\label{MatrixElements}
\end{align}
represent the matrix elements associated with the transition from the even ($e$) and odd ($o$) parity in-gap  states, respectively, to the excited (or extended) states of the spectrum. In general, these matrix elements are different.  A key question is whether this difference reveals information regarding the existence of MBSs. Clearly, in instances where the in-gap modes possess a finite energy, specifically $\epsilon_0 \neq 0$, the occupation of these modes can be discerned from the location of the absorption peak, which occurs at $\omega = \epsilon_n \pm \epsilon_0$. Nonetheless, a situation of greater relevance arises when these modes remain anchored at zero energy.  In this scenario, apart from the variation in strength, it is crucial that the condition $f(\epsilon_0)\neq1/2$ is satisfied, which signifies a departure from the ideal statistical mixture associated with degenerate energy levels. The difference is most pronounced when the parity is fixed within the  topological superconductor, as demonstrated previously for the case of a pristine topological wire  hosting MBSs at its ends 
\cite{Dmytruk_Phys.Rev.B.2023Microwave}.

In order to evaluate explicitly the susceptibility, we need to infer the quasiparticle distribution functions. The parity-constrained partition function reads \cite{Beenakker_Phys.Rev.Lett..2013FermionParity,Trif_Phys.Rev.B.2018Dynamic}
\begin{align}
    Z_{\mathcal{P}=\pm1}=\frac{Z_0}{2}\left[1+\mathcal{P}\prod_{\epsilon_n>0}\tanh\left(\frac{\beta\epsilon_n}{2}\right)\right]\,,
 \label{partition_parity}   
\end{align}
 where $Z_0=\prod_{\epsilon_n>0}2\cosh(\beta\epsilon_n/2)$ is the partition function without parity constraints, and $\beta=1/k_BT$, with $k_B$ representing the Boltzmann constant and $T$ denoting the temperature, respectively.  The function that describes the distribution for an energy state $\epsilon_n$ is expressed as $f(\epsilon_n) = 1/2 - (1/\beta)\,\partial\log Z_\mathcal{P}/\partial \epsilon_n$. In scenarios where there exists a zero-energy bound state, characterized by $\epsilon_0\rightarrow0$ in terms of $\beta\epsilon_0\ll1$, while ensuring $\epsilon_{n}$ remains nonzero for all $n>0$, the distribution function becomes
\begin{align}
f(\epsilon_n)=\left\{
    \begin{array}{cc}
f_0(\epsilon_n)\,, & \epsilon_n>0\\\\
\displaystyle{\frac{1}{2}}\left[1-\mathcal{P}\prod_{\epsilon_k>0}\left(1-2f_0(\epsilon_k)\right)\right]\,, & \epsilon_n=0\\
    \end{array}
    \right.\,,
\end{align}
where $f_0(\epsilon_n)=1/(e^{\beta\epsilon_n}+1)$ is the Fermi-Dirac distribution function. We see that at $T=0$, the parity is entirely dictated by the occupancy of the zero mode. However, at non-zero temperatures, this distinction diminishes due to the inequality $|\prod_{\epsilon_k>0}\left(1-2f_0(\epsilon_k)\right)|<1$. Moreover, as the system approaches the thermodynamic limit, meaning that the total number of modes becomes infinitely large, the zero mode's occupancy becomes independent of parity. Hence, we only focus on systems with finite sizes, as in the experimental implementations. 

For the parity-constrained susceptibility, we find
\begin{align}
\chi_{\mathcal{P}}(\omega)&=ig^2\bigg(\sum_{n\neq 0}\frac{|M_{n0}^t|^2[1-2f_0(\epsilon_n)]}{\omega-\epsilon_n+i\eta}\nonumber\\
&+\mathcal{P}\prod_{k\neq0}[1-2f_0(\epsilon_k)]\sum_{n\neq 0}\frac{|M_{n0}^d|^2}{\omega-\epsilon_n+i\eta}\bigg)\,,
\label{suscep_parity}
\end{align}
where $|M_{n0}^{t,d}|^2=|M_{n0}^{o}|^2\pm|M_{n0}^{e}|^2$  are the sum/difference in the transition strengths. The above equation allows us to define a visibility associated with the microwave absorption \cite{Shen_Phys.Rev.Research.2023Majoranamagnon}:
\begin{equation}
\begin{aligned}
    \nu(\omega,j_{c})&=  \frac{\text{Im}[\chi_o(\omega,j_c)] -\text{Im}[\chi_e(\omega,j_c)]}{\text{Im}[\chi_o(\omega,j_c)] +\text{Im}[\chi_e(\omega,j_c)]}\,,
\end{aligned}
\label{visibilityDef}
\end{equation}
where we employed the index $o$ ($e$) for odd (even) parity, denoted by $\mathcal{P}=1(-1)$, and emphasized the influence of the cavity coverage $j_c$, which is crucial for detection. 
To identify the criteria necessary for ensuring a finite value for visibility, it is beneficial to define:
\begin{align} 
    u_{0\sigma}^L(j)&=\frac{u_{0\sigma}(j)-v^*_{0\sigma}(j)}{i\sqrt{2}}\,;
    u_{0\sigma}^R(j)&=\frac{u_{0\sigma}(j)+v^*_{0\sigma}(j)}{\sqrt{2}}\,,
    \label{left_right}
\end{align}
which, for a pristine topological wire (in the nontrivial regime), labels the wavefunction of the MBS localized at the left ($L$) and right ($R$) ends, respectively \cite{Penaranda_Phys.Rev.B.2018Quantifying}. The corresponding probability densities are   
\begin{align}
    |\psi_{L,R}(j)|^2 = \sum_{\sigma=\uparrow,\downarrow} |u_\sigma^{L,R}(j)|^2\,.
\end{align}
The utility of the above definition is evident when the intensity transition difference is reformulated as follows: 
\begin{align}
    |M_{n0}^d|^2 = 2 \, {\rm Im}[(M_{n0}^{L})^{*} M_{n0}^R]\,,
    \label{Matrixelementsdifference}
\end{align}
where the definitions  $M_{n0}^{L,R}$ mirror the format in Eq.~\eqref{MatrixElements}, substituting $o$ and $e$ with $L$ and $R$, respectively. Thus, the visibility is nonzero only if the cavity couples simultaneously to both Majorana modes, providing a direct signature of their spatial nonlocality.

In what follows, we investigate in detail the features encoded by the visibility for MBSs, ABSs, and QMBSs, with the goal of identifying how their distinct behaviors can be used to distinguish among them.

\section{Main results}
\label{mainresults}
\begin{figure}[h]
  \centering
  \begin{tikzpicture}[remember picture, node distance=0cm]

    \node (suba) at (0,0) {
      \begin{minipage}{\linewidth}
        \centering
        \begin{tikzpicture}
          \node[anchor=south west,inner sep=0] (image) at (0,0) {
            \includegraphics[width=\linewidth, trim=2pt 0pt 0pt 0pt]{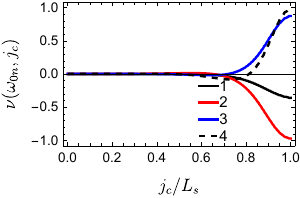}
          };
          \node[anchor=north west,font=\bfseries] at ([xshift=10pt,yshift=-3pt]image.north west) {(a)};
        \end{tikzpicture}
      \end{minipage}
    };

    \node[anchor=north west] (subb) at ([xshift=1.9cm,yshift=-0.1cm] suba.north west) {
      \begin{minipage}{0.41\linewidth}
        \centering
        \begin{tikzpicture}[x={(image.south east)},y={(image.north west)}]
          \node[anchor=south west] (ins1) at (0.22,0.14) {
            \includegraphics[width=\linewidth]{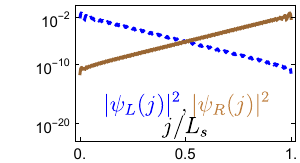}
          };
          \node[anchor=north west,font=\bfseries] at ([xshift=0pt,yshift=-3pt] ins1.north west) {(b)};
        \end{tikzpicture}
      \end{minipage}
    };

    \node[anchor=north west] (subc) at ([xshift=2cm,yshift=-1.9cm] suba.north west) {
      \begin{minipage}{0.4\linewidth}
        \centering
        \begin{tikzpicture}[x={(image.south east)},y={(image.north west)}]
          \node[anchor=south west] (ins2) at (0.22,0.135) {
            \includegraphics[width=\linewidth]{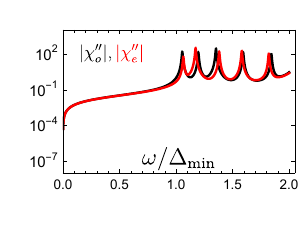}
          };
          \node[anchor=north west,font=\bfseries] at ([xshift=-1pt,yshift=-12pt] ins2.north west) {(c)};
        \end{tikzpicture}
      \end{minipage}
    };

  \end{tikzpicture}

  \caption{Microwave absorption of a pristine topological wire with MBSs. (a) The absorption visibility vs. the fraction of the wire that is coupled to cavity $j_c/L_s$ for several transitions from the MBS to excited states, ($n = 1, 2, 3, 4$, and $\omega_{0n}=\epsilon_n-\epsilon_0$). The visibility becomes nonzero only when the cavity overlaps with both MBSs. (b) The left (right) MBSs wavefunction probability amplitude $|\psi_{L(R)}(j)|^2$ along the wire, showing exponential localization at the left (right) edge. (c) The susceptibility for the odd (even) parity $|\chi''_{o(e)}(\omega)|$, for a nanowire that is fully coupled to the cavity ($j_c=L_s$) as a function of probe frequency $\omega$. The difference in the peak amplitude helps to distinguish the two parities. Here $\Delta_{\rm min}$ is the (minimal) SC gap for a wire with $L_s=500$ sites, while the values of all model parameters are listed in Table~\ref{parametertable}. 
  }
  \label{PristineMajorana_VX_WF_VW}
\end{figure}

\begin{figure*} 
    \centering

    \begin{minipage}[t]{0.32\linewidth}
        \centering
        \begin{tikzpicture}
            \node[anchor=south west, inner sep=0] (image) at (0,0) {
                \includegraphics[width=\linewidth]{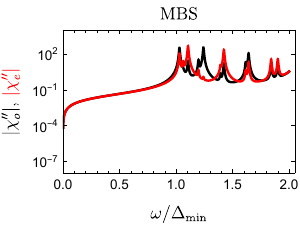}
            };
            \begin{scope}[x={(image.south east)}, y={(image.north west)}]
                \node[anchor=south west] at (0.44,0.23) {
                    \includegraphics[width=0.51\linewidth]{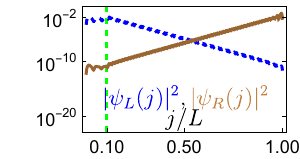}
                };
                \node[anchor=north west, font=\small\bfseries] at (0.0,1.02) {(a)};
            \end{scope}
        \end{tikzpicture}
    \end{minipage}
    \hfill
    \begin{minipage}[t]{0.32\linewidth}
        \centering
        \begin{tikzpicture}
            \node[anchor=south west, inner sep=0] (image) at (0,0) {
                \includegraphics[width=\linewidth]{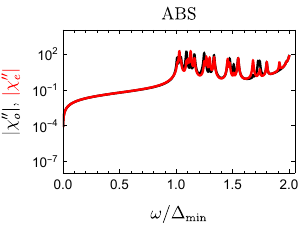}
            };
            \begin{scope}[x={(image.south east)}, y={(image.north west)}]
                \node[anchor=south west] at (0.44,0.23) {
                    \includegraphics[width=0.51\linewidth]{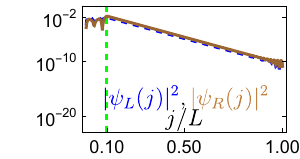}
                };
                \node[anchor=north west, font=\small\bfseries] at (0.03,1.02) {(b)};
            \end{scope}
        \end{tikzpicture}
    \end{minipage}
    \hfill
    \begin{minipage}[t]{0.32\linewidth}
        \centering
        \begin{tikzpicture}
            \node[anchor=south west, inner sep=0] (image) at (0,0) {
                \includegraphics[width=\linewidth]{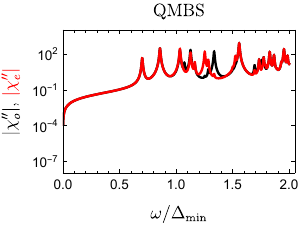}
            };
            \begin{scope}[x={(image.south east)}, y={(image.north west)}]
                \node[anchor=south west] at (0.44,0.23) {
                    \includegraphics[width=0.51\linewidth]{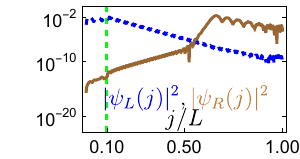}
                };
                \node[anchor=north west, font=\small\bfseries] at (0.03,1.02) {(c)};
            \end{scope}
        \end{tikzpicture}
    \end{minipage}


    \begin{minipage}[t]{0.32\linewidth}
        \centering
        \begin{tikzpicture}
            \node[anchor=south west, inner sep=0] (image) at (0,0) {
                \includegraphics[width=\linewidth]{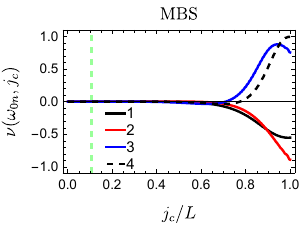}
            };
            \begin{scope}[x={(image.south east)}, y={(image.north west)}]
                \node[anchor=north west, font=\small\bfseries] at (0.03,1.02) {(d)};
            \end{scope}
        \end{tikzpicture}
    \end{minipage}
    \hfill
    \begin{minipage}[t]{0.32\linewidth}
        \centering
        \begin{tikzpicture}
            \node[anchor=south west, inner sep=0] (image) at (0,0) {
                \includegraphics[width=\linewidth]{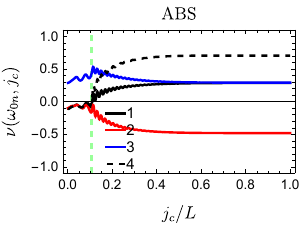}
            };
            \begin{scope}[x={(image.south east)}, y={(image.north west)}]
                \node[anchor=north west, font=\small\bfseries] at (0.03,1.02) {(e)};
            \end{scope}
        \end{tikzpicture}
    \end{minipage}
    \hfill
    \begin{minipage}[t]{0.32\linewidth}
        \centering
        \begin{tikzpicture}
            \node[anchor=south west, inner sep=0] (image) at (0,0) {
                \includegraphics[width=\linewidth]{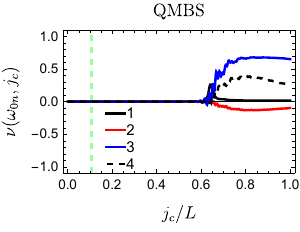}
            };
            \begin{scope}[x={(image.south east)}, y={(image.north west)}]
                \node[anchor=north west, font=\small\bfseries] at (0.03,1.02) {(f)};
            \end{scope}
        \end{tikzpicture}
    \end{minipage}\\
    %
 \begin{minipage}[t]{0.32\linewidth}
        \centering
        \begin{tikzpicture}
            \node[anchor=south west, inner sep=0, outer sep=0] (image) at (0,0) {
                \includegraphics[width=\linewidth]{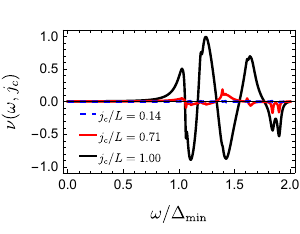}
            };
            \begin{scope}[x={(image.south east)}, y={(image.north west)}]
                \node[anchor=north west, font=\small\bfseries] at (0.03,1.02) {(g)};
            \end{scope}
        \end{tikzpicture}
    \end{minipage}
     \hfill
    \begin{minipage}[t]{0.32\linewidth}
        \centering
        \begin{tikzpicture}
            \node[anchor=south west, inner sep=0, outer sep=0] (image) at (0,0) {
                \includegraphics[width=\linewidth]{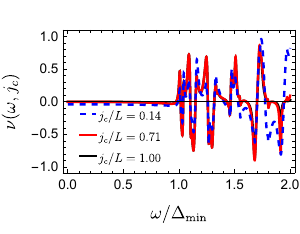}
            };
            \begin{scope}[x={(image.south east)}, y={(image.north west)}]
                \node[anchor=north west, font=\small\bfseries] at (0.03,1.02) {(h)};
            \end{scope}
        \end{tikzpicture}
    \end{minipage}
    \hfill
    \begin{minipage}[t]{0.32\linewidth}
        \centering
        \begin{tikzpicture}
            \node[anchor=south west, inner sep=0, outer sep=0] (image) at (0,0) {
                \includegraphics[width=\linewidth]{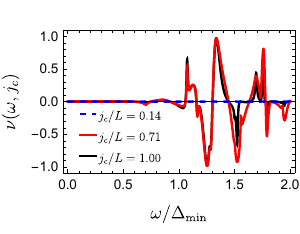}
            };
            \begin{scope}[x={(image.south east)}, y={(image.north west)}]
                \node[anchor=north west, font=\small\bfseries] at (0.03,1.02) {(i)};
            \end{scope}
        \end{tikzpicture}
    \end{minipage} 
    \caption{Microwave absorption of a topological wire with an adjacent QD (vertical green dashed line marks the QD–SC interface). [(a)--(c)] Imaginary part of the susceptibility $|\chi_{o(e)}''(\omega,j_c)|$ for the odd (even) parity as a function of the frequency $\omega$  when the wire is fully covered by the cavity.  The insets show the probability left (right) amplitudes $|\psi_{L(R)}(j)|^2$ corresponding to the zero-energy states as function of the position $j$ in the wire for MBSs, ABSs and QMBSs, respectively. [(d)--(f)] Visibility $\nu(\omega_{0n},j_{c})$ as function of the last site that couples to the cavity, $j_c$,  for transitions between the zero-energy state and the excited states $n=1,2,3,4$ for MBSs, ABSs and QMBSs cases, respectively. Unlike the ABSs and QMBSs, the MBSs visibility vanishes unless the cavity couples to both SC edges, a signature for their nonlocality. Figures [(g)--(i)] show visibility $\nu(\omega, j_c)$ as a function of the probe frequency $\omega$ for different  fractions $j_c/L = 0.14, 0.71, 1$.  (g) MBSs visibility, which becomes nonzero only when the cavity couples to both Majorana end modes at resonant $\omega$. (h) ABSs localized near the interface yield nonzero visibility even when the coupling to the cavity only occurs near the left edge (i) The QMBSs  exhibit nonzero visibility once the cavity covers both of them, which happens for $j_c/L>0.64$ for the depicted setup. The parameters utilized are presented in Table \ref{parametertable} for each case. 
    }
    \label{cleancases}
\end{figure*}

This section summarizes our core findings. We benchmark the microwave absorption visibility using a pristine topological nanowire without a QD. In this case, Fig.~\ref{PristineMajorana_VX_WF_VW}(a) shows that the visibility becomes nonzero only when the cavity overlaps with both MBSs localized at the ends of the wire [see Fig.~\ref{PristineMajorana_VX_WF_VW}(b) for the plots of $|\psi_{L,R}(j)|^2$]. Then we  calculate the visibility for scenarios in the presence of QD where the emerging zero-energy excitations are MBSs, ABSs, and QMBSs, respectively. Our results indicate that the unique visibility profiles of these cases can be instrumental in differentiating MBSs from ABSs and QMBSs. As illustrated in the inset of Fig.~\ref{cleancases}(a), the MBSs probability amplitude exhibits pronounced peaks at the boundaries of the superconductor. The microwave absorption visibility, which depends on the position $j_c$ of the last site interacting with the cavity, only takes on values other than zero when the cavity covers both edges of the SC (since only in that case $|M_{n0}^{d}|\neq0$), as seen in Fig.~\ref{cleancases}(d) and Fig.~\ref{cleancases}(g). Consequently, the visibility effectively illustrates the nonlocal characteristic of Majorana's spatial separation. In contrast, as depicted in the inset of Fig.~\ref{cleancases}(b), the Andreev state is localized in proximity to the interface with the normal part (i.e., the quantum dot), leading to a visibility in  Fig.~\ref{cleancases}(e) and  Fig.~\ref{cleancases}(h) that saturates as more sites in the SC region couple to the cavity, as there are no more in-gap states to alter the transition's strength.  In Fig.~\ref{cleancases}(f) and Fig.~\ref{cleancases}(i),
we depict the quasi-Majorana visibility reaching its maximum when the cavity engages with both QMBS wavefunction peaks [see inset of Fig.~\ref{cleancases}(c)] at the boundaries of the topological phase. This behavior contrasts with the true MBS scenario, where their localization consistently occurs at the topological superconductor's edges.

\subsection{Visibility of MBSs in a nanowire without a QD}
To benchmark our visibility calculations, we first consider a  superconducting Rashba wire in a magnetic field and in the absence of the  QD. Moreover, we assume that the system is tuned into the topological nontrivial regime. In this case, a pair of MBSs appears at its edges. Their spatial extension, denoted as $\xi_{\rm MBS}\propto v_F/\Delta_{{\rm min}}$, is assumed to be significantly smaller than the length $L_s$ of the system ($\xi_{\rm MBS}/L_s\approx0.1$ in Fig.~3), resulting in negligible overlap between them. In Fig.~\ref{PristineMajorana_VX_WF_VW}(a), we show the visibility of the microwave absorption, \(\nu(\omega,j_c)\), as a function of the cut-off site index \(j_c\) up to which the cavity field is coupled to the wire. We focus on the  transitions between the MBS (\(n=0\)) and the first four excited states  (\(n=1,2,3,4\)), with transition frequencies \(\omega\equiv\omega_{0n} = \epsilon_{n}-\epsilon_{0}\). In this case, the visibility 
\begin{equation}
   \nu(\omega_{0n},j_c) \propto \bigl|M_{m0}^{d}(j_c)\bigr|^2 
\end{equation}
vanishes unless the upper limit of the summation $j_c$ is such that the sum includes both spatially separated Majorana modes $|\psi_{L(R)}(j)|^2$ at the SC edges [shown in Fig.~\ref{PristineMajorana_VX_WF_VW}(b)]. Moreover, we see that the visibility for a given transition is associated with a particular parity that can help distinguish between odd and even parities. For example, the transition to $n=1,2$ is associated with even parity, while $n=3,4$ can be associated with odd parity. Such behavior, for the pristine wire, originates from the presence of inversion symmetry, as discussed in detail in Ref.~\cite{Shen_Phys.Rev.Research.2023Majoranamagnon}.

Finally, for completeness, in Fig.~\ref{PristineMajorana_VX_WF_VW}(c) we depict the imaginary part of the odd (even) parity susceptibility pertaining to the transition between the zero-mode and the first excited state, $|\chi_{o(e)}''(\omega)|$, for a wire that is fully covered by the cavity, $j_c=L_s$. We see that the two parity sectors exhibit distinct resonance peaks, although they share the same resonance frequency. We stress that when the cavity does not fully couple to the wire (not shown here, see, for example, Ref.~\cite{Dmytruk_Phys.Rev.B.2023Microwave}), the susceptibility curves for the two parities are indistinguishable, resulting in a vanishing visibility.

Next, we consider the presence of an adjacent QD to the superconducting Rashba wire and examine how this coupling modifies the visibility features for the various emergent zero-energy modes.

\subsection{Visibility of MBSs with a QD}

We assume that the wire parameters are such that the SC segment is in the nontrivial regime (for the parameters, see Table~\ref{parametertable}).  
The probability amplitude $|\psi_{L(R)}(j)|^2$ for the left (right) zero-energy state ($\epsilon_0/\Delta_{\rm min}=3.25\times 10^{-5}$) is displayed on a logarithmic scale in the inset of Fig.~\ref{cleancases}(a) to emphasize their exponential localization near the superconductor edges. We see that $|\psi_{L}(j)|^2$ exhibits a reduced amplitude compared to the pristine case, which is due to its partial leakage into the QD. 

The main panel of Fig.~\ref{cleancases}(a) displays $|\chi_{o,e}''(\omega)|$ as a function of the probing frequency $\omega$ when the cavity completely covers the wire (see Fig.~\ref{ImChi_W_AllCases_CleanQD} in the Appendix for plots of $|\chi_{o,e}''(\omega)|$ for other covering fractions). The two parity sectors exhibit the same positions of the absorption peaks but different amplitudes, just as in the case of the pristine Majorana wire. This behavior extends the susceptibility analysis presented in Ref.~\cite{Dmytruk_Phys.Rev.B.2023Microwave}, allowing one to distinguish the two parities by scanning the probe frequency in the presence of a QD. The difference in the peak amplitudes for odd and even parities as a function of $\omega$ depends on the fraction of cavity covering lengths $j_c/L$, as shown in Fig.~\ref{cleancases}(g). When the cavity extends only to $j_c/L = 0.14$—which lies close to the QD-SC interface at $L_{d}/L = 0.1$—the visibility remains zero across all resonant transitions (hence, it cannot distinguish parities), since the cavity couples solely to the left MBS.  The same argument applies when the cavity covers up to $j_c/L = 0.71$. However, once the cavity covers the entire wire ($j_c/L = 1$), the visibility becomes nonzero at each resonant transition. 

In Fig.~\ref{cleancases}(d) we show the visibility $\nu(\omega_{0n},j_c)$ as a function of $j_c/L$ for transitions from the zero‐energy Majorana mode to the first four excited states $n=1,2,3,4$.  All transitions exhibit a nonzero visibility when the cavity couples to both MBSs due to the nature of the matrix element in Eq.~\ref{Matrixelementsdifference}, similar to the pristine case. Therefore, the Majorana nonlocality is probed by requiring that the cavity overlap both end modes even in the presence of QD.

\subsection{Visibility of zero-energy ABSs}
\begin{figure*}
    \centering

    \begin{minipage}[t]{0.32\linewidth}
        \centering
        \begin{tikzpicture}
            \node[anchor=south west, inner sep=0, outer sep=0] (image) at (0,0) {
                \includegraphics[width=\linewidth]{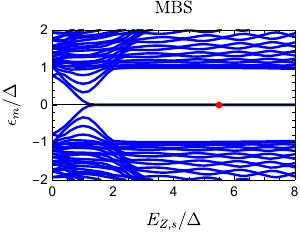}
            };
            \begin{scope}[x={(image.south east)}, y={(image.north west)}]
                \node[anchor=north west, font=\small\bfseries] at (0.03,1.02) {(a)};
            \end{scope}
        \end{tikzpicture}
    \end{minipage}
     \hfill
    \begin{minipage}[t]{0.32\linewidth}
        \centering
        \begin{tikzpicture}
            \node[anchor=south west, inner sep=0, outer sep=0] (image) at (0,0) {
                \includegraphics[width=\linewidth]{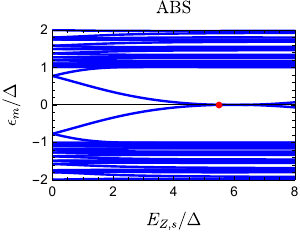}
            };
            \begin{scope}[x={(image.south east)}, y={(image.north west)}]
                \node[anchor=north west, font=\small\bfseries] at (0.03,1.02) {(b)};
            \end{scope}
        \end{tikzpicture}
    \end{minipage}
    \hfill
    \begin{minipage}[t]{0.32\linewidth}
        \centering
        \begin{tikzpicture}
            \node[anchor=south west, inner sep=0, outer sep=0] (image) at (0,0) {
                \includegraphics[width=\linewidth]{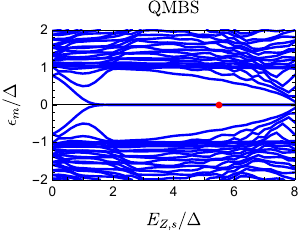}
            };
            \begin{scope}[x={(image.south east)}, y={(image.north west)}]
                \node[anchor=north west, font=\small\bfseries] at (0.03,1.02) {(c)};
            \end{scope}
        \end{tikzpicture}
    \end{minipage}
    \caption{Spectrum of the electronic system as a function of the Zeeman energy, normalized by the proximity-induced gap $\Delta$. (a) Topological superconducting wire coupled to a QD displaying MBSs at its endpoints.  In (b), the system is in a trivial regime characterized by accidental zero-energy ABSs localised at the QD-SC interface in some region of Zeeman energy.  (c) System supports  QMBSs arising from an inhomogeneous chemical potential. Each schematic includes red dots marking the positions where susceptibility and visibility analyses are conducted.  
    }
    \label{EnergyWithZeeman}
\end{figure*}
Zero-energy ABSs can arise due to different reasons; for example, variation in chemical potential \cite{Kells_Phys.Rev.B.2012Nearzeroenergy,Liu_Phys.Rev.B.2018Distinguishing, Vuik_SciPostPhys..2019Reproducing}, interband effects \cite{Woods_Phys.Rev.B.2019Zeroenergy, Chen_Phys.Rev.Lett..2019Ubiquitous}, or the presence of a QD~\cite{Reeg_Phys.Rev.B.2018Zeroenergy,Liu_Phys.Rev.B.2017Andreev,Ptok_Phys.Rev.B.2017Controlling}. As Ref.~\cite{Reeg_Phys.Rev.B.2018Zeroenergy} already investigated the conditions for the emergence of  trivial zero-energy ABSs that mimic MBSs in a nanowire with a QD akin to our model, we follow their setup and set the SOI coupling strength $\alpha_s = 0$ to ensure that the superconducting region remains in the trivial phase. Moreover, similar to Ref.~\cite{Reeg_Phys.Rev.B.2018Zeroenergy}, we choose a Zeeman energy that lies within a range of Zeeman energies that give zero-energy ABSs [see Fig.\ref{EnergyWithZeeman}(b)], with the rest of the parameters as in Table~\ref{parametertable}. With this choice of parameters, we obtain an ABS with near-zero energy  $\epsilon_0/\Delta_{\rm min}=2.76\times10^{-4}$. The left  (right) wave‐function probability amplitude $|\psi_{L,R}(j)|^2$ is plotted as a function of position in the inset of Fig.~\ref{cleancases}(b), clearly showing that both $|\psi_{L}(j)|^2$ and $|\psi_{R}(j)|^2$  are localized near the QD–SC interface, overlapping strongly. Furthermore, they decay exponentially away from the  interface into the SC, similarly to the MBSs.

The main panel of Fig.~\ref{cleancases}(b) depicts the imaginary part of the susceptibility, $|\chi_{o(e)}''(\omega,j_c)|$, plotted against the normalized probe frequency $\omega/\Delta_{\rm min}$ when the cavity completely covers the entire wire (see Fig.~\ref{ImChi_W_AllCases_CleanQD} in the Appendix for plots of $|\chi_{o,e}''(\omega)|$ for other covering fractions). As in the MBS case, the two parity sectors produce peaks with different amplitudes, enabling parity discrimination by sweeping $\omega$.  However, the peak structure remains distinct even when the cavity coverage fraction $j_c/L$ is reduced, which conforms with the localization of
ABS at the interface and is in stark contrast to the MBSs.  
In Fig.~\ref{cleancases}(e), we plot the visibility $\nu(\omega_{0n},j_c)$ as a function of $j_c/L$ for transitions from the zero‐energy ABS to excited states labeled by $n = 1,2,3,4$.  For each transition, the visibility attains a fixed sign—hence a fixed parity—even before the cavity spans the entire wire. This occurs because the matrix element in Eq.~\ref{Matrixelementsdifference} saturates once the cavity covers the region around the interface. Figure~\ref{cleancases}(h) shows the visibility as a function of $\omega$ for various cavity covering fractions. Even for a cavity fraction $j_c/L = 0.14$, the visibility becomes nonzero, as the cavity already couples to the peaks near the QD–SC interface.

\subsection{Visibility of zero-energy QMBSs}\label{VisibilityofzeroenergyQMBSs}
We define QMBSs as zero‐energy states that arise from unintended variations in the chemical potential due to experimental imperfections.  For example, attempting to impose a steplike chemical potential profile can yield a smoother spatial variation in practice, since gate voltages may not be perfectly sharp. Such smooth potentials can support zero‐energy states that exhibit exponential localization similar to true MBSs \cite{Vuik_SciPostPhys..2019Reproducing,Kells_Phys.Rev.B.2012Nearzeroenergy}.
\begin{figure}
    \centering
    \includegraphics[width=0.7\linewidth,
    trim={4.5cm 2.7cm 5cm 2cm}, clip]{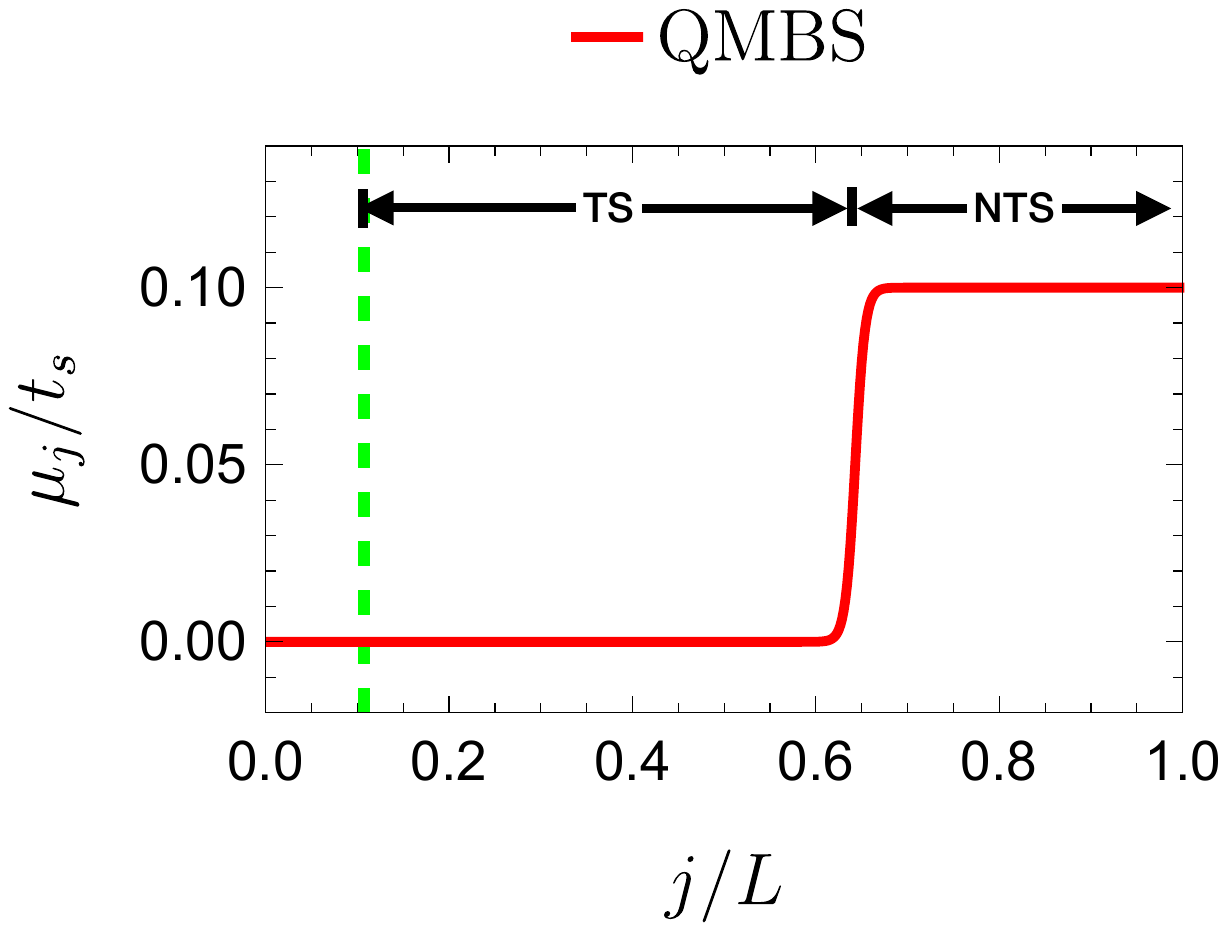}
    \caption{Plot of the chemical-potential profile in Eq.~\eqref{ChemcialPotentialForArxivQMBSProfile}, based on the parameters in Table~\ref{parametertable}, indicating the topological superconducting (TS) and nontopological superconducting (NTS) region. In this configuration, the TS region converts to an NTS region at $j_\xi / L \approx 0.64$, yielding a minimum energy of $\epsilon_0 / \Delta_{\min} = 9.52 \times 10^{-6}$. Although this corresponds to one realization of a QMBS (e.g., with $j_\xi / L$ located at the QD-SC interface), Appendix~\ref{NewQMBSAppendix} presents an alternative setup where the TS region is shorter than the NTS one, which increases $\epsilon_0 / \Delta_{\min}$ due to stronger wavefunction overlap.
    }
    \label{QMBSMuProfile}
\end{figure}

Different smooth-potential profiles can generate various types of QMBSs~\cite{Kells_Phys.Rev.B.2012Nearzeroenergy,Prada_Phys.Rev.B.2012Transport,Penaranda_Phys.Rev.B.2018Quantifying,Vuik_SciPostPhys..2019Reproducing}, but we focus on two representative configurations to study their influence on visibility. In the first configuration analyzed in the main text, the smooth potential produces a topological region that is longer than the trivial portion of the superconducting wire. The second configuration, detailed in Appendix~\ref{NewQMBSAppendix}, is the opposite: The topological region is shorter than the trivial one, which enhances the overlap between the QMBS peaks and yields a higher lowest-energy state than in the first case. We concentrate on the first configuration here because its lowest-energy state lies very close to zero, akin to the MBS scenario, making it particularly suitable for examining visibility.

To model both scenarios, we assume a smooth chemical potential profile along the SC region.  Specifically, for sites $j > L_{d}$, we consider a chemical potential of the form
\begin{equation}
\mu_j \;=\; \mu_0 \left[\,1 + \tanh\!\Bigl(\displaystyle{\tfrac{j - j_{\zeta}}{\zeta_0}}\Bigr)\right],
\label{ChemcialPotentialForArxivQMBSProfile}
\end{equation}
where $\mu_0$ is the maximum height of the smooth potential, $j_\zeta$ marks the position of the topological–trivial interface, and $\zeta_0$ controls the spatial width. For the plots corresponding to the first case, we chose $j_\zeta/L = 0.64$ and $\zeta_0/L = 0.01$, resulting in the chemical potential profile illustrated in Fig.~\ref{QMBSMuProfile}. Consequently, sites $j \in [L_d,\, j_\zeta]$ lie in the topological phase, while sites $j > j_\zeta$ remain trivial. This profile yields two QMBS peaks in $|\psi_{L(R)}(j)|^2$ at the boundaries of the effective topological region [inset of Fig.~4(c)], and supports a mode with energy pinned close to zero, $\epsilon_0/\Delta_{\min} = 9.52 \times 10^{-6}$ (see Table~\ref{parametertable}).

Figure~\ref{cleancases}(c) shows the imaginary part of the susceptibility, $|\chi''_{o(e)}(\omega)|$, as a function of $\omega/\Delta_{\min}$ when the cavity fully covers the wire (see Fig.~\ref{ImChi_W_AllCases_CleanQD} in the Appendix for $|\chi''_{o,e}(\omega)|$ at other coverage fractions). As in the MBS and ABS cases, even and odd parities exhibit distinct resonance features, enabling parity readout by sweeping $\omega$. On the other hand, Fig.~\ref{cleancases}(f) displays $\nu(\omega_{0n}, j_c)$ versus $j_c/L$ for transitions from the zero-energy QMBS to the excited states $\epsilon_n$. Nonzero visibility first appears precisely when the cavity field spans both localized QMBS wave-function lobes at the edges of the effective topological region. In contrast to true MBSs, QMBSs can therefore exhibit finite visibility before the cavity covers the entire SC: The onset is set by the extent of the \emph{effective} topological segment defined by the smooth potential, not by the physical ends of the wire.

Finally, Fig.~\ref{cleancases}(i) shows visibility $\nu(\omega,j_c)$ as a function of $\omega$ for different cavity covering fraction $j_c/L$. When the cavity couples only to sites just past the interface (e.g., $j_c/L = 0.14$), the visibility remains zero even as $\omega$ crosses resonances. When the coupling extends to $j_c/L = 0.71$, the cavity overlaps both QMBS lobes (near $j_c/L \approx 0.14$ and $j_\zeta/L \approx 0.64$), and a finite visibility emerges at the corresponding resonant frequencies. For full coverage ($j_c/L = 1$), sharp visibility peaks appear whenever $\omega$ matches a transition out of the zero-energy QMBS, clearly distinguishing the parities.

In summary, QMBSs mimic MBSs spectrally, but their visibility reveals only partial nonlocality: Finite visibility arises once the cavity overlaps both QMBS lobes inside the SC, without requiring full-wire coverage. This provides a clear diagnostic distinction from genuine MBSs.

\section{Effect of chemical potential disorder}\label{EffectOfDisorder}
\begin{figure*}[t]
    \centering

    \begin{minipage}[t]{0.32\linewidth}
        \centering
        \begin{tikzpicture}
            \node[anchor=south west, inner sep=0, outer sep=0] (image) at (0,0) {
                \includegraphics[width=\linewidth]{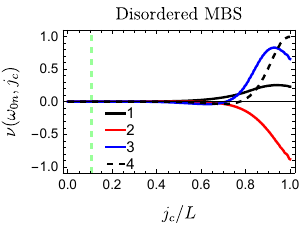}
            };
            \begin{scope}[x={(image.south east)}, y={(image.north west)}]
                \node[anchor=north west, font=\small\bfseries] at (0.03,1.02) {(a)};
            \end{scope}
        \end{tikzpicture}
    \end{minipage}
    \hfill
    \begin{minipage}[t]{0.32\linewidth}
        \centering
        \begin{tikzpicture}
            \node[anchor=south west, inner sep=0, outer sep=0] (image) at (0,0) {
                \includegraphics[width=\linewidth]{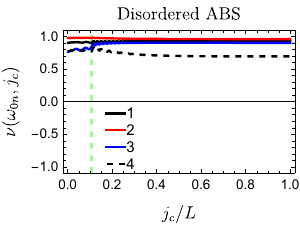}
            };
            \begin{scope}[x={(image.south east)}, y={(image.north west)}]
                \node[anchor=north west, font=\small\bfseries] at (0.03,1.02) {(b)};
            \end{scope}
        \end{tikzpicture}
    \end{minipage}
    \hfill
    \begin{minipage}[t]{0.32\linewidth}
        \centering
        \begin{tikzpicture}
            \node[anchor=south west, inner sep=0, outer sep=0] (image) at (0,0) {
                \includegraphics[width=\linewidth]{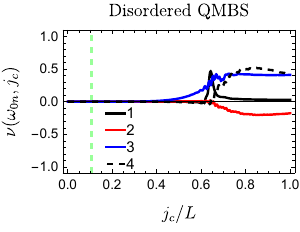}
            };
            \begin{scope}[x={(image.south east)}, y={(image.north west)}]
                \node[anchor=north west, font=\small\bfseries] at (0.03,1.02) {(c)};
            \end{scope}
        \end{tikzpicture}
    \end{minipage}
    \caption{The visibility in the presence of a Gaussian disorder in the chemical potential for (a) MBSs, (b) ABSs, and (c) QMBSs, with $\sigma_{\rm dis} = \Delta/2$ in the SC for one realization (see Table~\ref{parametertable} for the parameter values and the vertical green dashed line marks the QD–SC interface). The  nonlocality of the MBSs, rooted in its topological origin, helps maintain its visibility characteristics, allowing parity distinction only when the cavity couples the full wire, even with potential disorder. In contrast, the ABSs and QMBSs, which lack such protection, show altered behavior compared to the clean case. 
    }
    \label{disordercases}
\end{figure*}

In this section, we investigate how spatial fluctuations in the chemical potential affect the visibility of the MBSs, ABSs, and QMBSs, respectively.  In a realistic device, the electrostatic landscape is never perfectly uniform; gate voltages and material inhomogeneities introduce random variations in the local chemical potential. To model this, we add a Gaussian disorder  potential $\delta\mu_j$ on each lattice site $j$ of the SC wire, drawn from a normal distribution with zero mean and a standard deviation $\sigma_{\rm dis} = \Delta/{2}$, where $\Delta$ is the induced superconducting gap (see Table~\ref{parametertable} for numerical values). 

Disorder in the chemical potential modulates the gap and the shape of both the localized and extended modes wave functions. In a topological SC, moderate disorder that does not close the bulk gap preserves the MBSs via topological protection.  In contrast, trivial ABSs and QMBSs lack global topological protection, so their energies and spatial profiles can be strongly affected even by weak disorder. In order to align with experimental conditions (specifically, low temperatures, gate-tunable homogeneities, and device-specific features), we consider individual disorder instances rather than an average across all potential realizations \cite{Pan_Phys.Rev.B.2021Threeterminal,Pan_Phys.Rev.B.2021Disorder}.

In Fig.~\ref{disordercases}(a), we depict the visibility $\nu(\omega_{0n},j_c)$ for a disordered wire that supports MBS as a function of $j_c/L$ for the transitions between the zero‐energy MBS and the excited states labeled by $n=1,2,3,4$.  Despite the random fluctuations in $\mu_j$, the MBSs remain pinned at zero energy ($\epsilon_0/ \Delta_{\rm min}=3.22\times 10^{-5}$) because the topological gap remains open throughout the wire. Consequently, the visibility retains its characteristic nonlocal behavior: It remains zero until the cavity overlaps with both MBSs, thereby distinguishing parities only when the wire is fully covered by the cavity field (or couples to the end regions of the wire). 

We analyzed the ABS scenario under conditions where disorder occurs in the superconducting part, using the parameters listed in Table~\ref{parametertable}. Spatial fluctuations of the chemical potential shift the ABS energy away from zero ($\epsilon_0/\Delta_{\rm min}=2.51\times 10^{-2}$) compared to its clean case  because, unlike MBS, the ABS is not protected by a topological gap and relies on fine‐tuned conditions (e.g., the chemical potential and Zeeman field) for zero-energy pinning.  Figure~\ref{disordercases}(b) shows the visibility $\nu(\omega_{0n},j_c)$ as a function of $j_c/L$, and similar to the clean case, we see that the two parities are distinguished before the cavity couples to the right SC edge, making it distinct from MBSs.

Even in the presence of moderate disorder, the QMBSs remain pinned near zero energy ($\epsilon_0/\Delta_{\rm min}=3.91 \times 10^{-5}$). Hence, they are capable of mimicking the MBSs in the disordered limit as well, as shown in Fig.~\ref{disordercases}(c). We have checked that this remains true for other disorder realizations; see Appendix \ref{Appendix_Disorder_2} for another realization. Nevertheless, for all plots, the magnitude of the parity-dependent susceptibility (and consequently the visibility) changes drastically even with weak disorder. 
This behavior arises because these quantities involve the wave functions of extended states, which are more susceptible to disorder than in-gap states, as the energy separation between different states decreases as $\propto 1/L^{2}$ with system size \cite{Dmytruk_Phys.Rev.B.2023Microwave}.

\section{Effect of a tunneling barrier}\label{barriersection}
\begin{figure*}[t]
    \centering
    \begin{minipage}[t]{0.32\linewidth}
        \centering
        \begin{tikzpicture}
            \node[anchor=south west, inner sep=0, outer sep=0] (image) at (0,0) {
                \includegraphics[width=\linewidth]{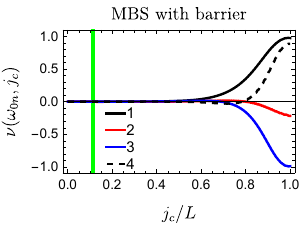}
            };
            \begin{scope}[x={(image.south east)}, y={(image.north west)}]
                \node[anchor=north west, font=\small\bfseries] at (0.03,1.02) {(a)};
            \end{scope}
        \end{tikzpicture}
    \end{minipage}
    \hfill
    \begin{minipage}[t]{0.32\linewidth}
        \centering
        \begin{tikzpicture}
            \node[anchor=south west, inner sep=0, outer sep=0] (image) at (0,0) {
                \includegraphics[width=\linewidth
                ]{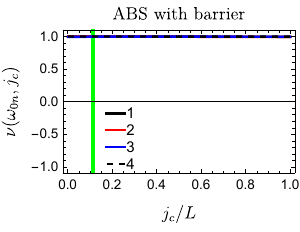}
            };
            \begin{scope}[x={(image.south east)}, y={(image.north west)}]
                \node[anchor=north west, font=\small\bfseries] at (0.03,1.02) {(b)};
            \end{scope}
        \end{tikzpicture}
    \end{minipage}
    \hfill
    \begin{minipage}[t]{0.32\linewidth}
        \centering
        \begin{tikzpicture}
            \node[anchor=south west, inner sep=0, outer sep=0] (image) at (0,0) {
                \includegraphics[width=\linewidth]{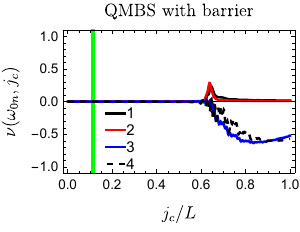}
            };
            \begin{scope}[x={(image.south east)}, y={(image.north west)}]
                \node[anchor=north west, font=\small\bfseries] at (0.03,1.02) {(c)};
            \end{scope}
        \end{tikzpicture}
    \end{minipage}
    \begin{minipage}[t]{0.32\linewidth}
        \centering
        \begin{tikzpicture}
            \node[anchor=south west, inner sep=0, outer sep=0] (image) at (0,0) {
                \includegraphics[width=\linewidth]{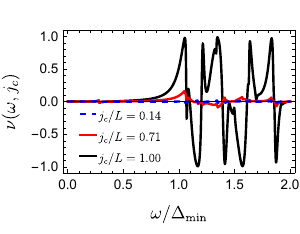}
            };
            \begin{scope}[x={(image.south east)}, y={(image.north west)}]
                \node[anchor=north west, font=\small\bfseries] at (0.03,1.02) {(d)};
            \end{scope}
        \end{tikzpicture}
    \end{minipage}
    \hfill
    \begin{minipage}[t]{0.32\linewidth}
        \centering
        \begin{tikzpicture}
            \node[anchor=south west, inner sep=0, outer sep=0] (image) at (0,0) {
                \includegraphics[width=\linewidth]{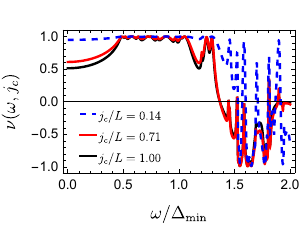}
            };
            \begin{scope}[x={(image.south east)}, y={(image.north west)}]
                \node[anchor=north west, font=\small\bfseries] at (0.03,1.02) {(e)};
            \end{scope}
        \end{tikzpicture}
    \end{minipage}
    \hfill
    \begin{minipage}[t]{0.32\linewidth}
        \centering
        \begin{tikzpicture}
            \node[anchor=south west, inner sep=0, outer sep=0] (image) at (0,0) {
                \includegraphics[width=\linewidth]{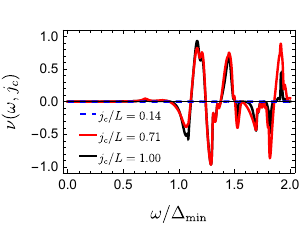}
            };
            \begin{scope}[x={(image.south east)}, y={(image.north west)}]
                \node[anchor=north west, font=\small\bfseries] at (0.03,1.02) {(f)};
            \end{scope}
        \end{tikzpicture}
    \end{minipage}

    \caption{
    Visibility in the presence of a barrier with a height of $\mu_{b}/t_{s}=3.5$, spanning the interval $j\in [L_{d}, L_{d}+L_{b}]$, where $L_{b}=10$ denotes its length (green shaded area marks the barrier region starting from the QD–SC interface). In (a), MBS visibility becomes nonzero only when the cavity couples to the full wire --- a key distinguishing feature. Panel (d) shows visibility versus $\omega$, again confirming that full-wire coupling is needed for MBS parity resolution. In (b), the ABS visibility distinguishes the two parities even without covering the full wire because the barrier shifts the zero energy leading to different peaks for the imaginary part of susceptibility for the two parities (see Fig.~\ref{Appendix_Barrier_ImChi_W} in Appendix \ref{ImChi_W_AllCases}). In (e), ABS visibility confirms parity distinction before full-wire coupling. In (c), QMBS parity is resolved when the cavity couples to the local topological edges, and (f) shows corresponding visibility versus $\omega$, which is nonzero only when cavity couples to local topological edges. Thus, full-wire coupling as a requirement for nonzero visibility is a distinctive MBS signature, setting it apart from ABS and QMBS. All other parameters are shown in Table~\ref{parametertable}.
    }
    \label{barriercases}
\end{figure*}

In this  section, we explore how a tunnel barrier influences the visibility of microwave absorption. By adjusting the voltages of adjacent gates, both the height and width of the barrier can be controlled. In our calculations, we assume the barrier is positioned within the wire at sites $j \in [L_{d}, L_{d}+L_b]$. Here we have chosen $L_{b}/L = 0.01$, and the barrier height is set as $\mu_{b}/t_{s}=3.5$. We refer to Table~\ref{parametertable} for the additional parameters. We stress that although different choices of tunnel parameters will alter the quantitative results, the qualitative findings will remain the same.


With the barrier in place, the lowest‐energy MBS remains pinned at $\epsilon_0 /\Delta_{\rm min}= 7.04\times10^{-5}$. In Fig.~\ref{barriercases}(a), we show the visibility $\nu(\omega_{0n},j_c)$ for MBSs as a function of the cavity coverage fraction $j_c/L$ for the four lowest-energy transitions ($n = 1,2,3,4$). Similarly to the case without a barrier, the visibility remains zero until the cavity couples to both 
Majorana end modes, at which point the visibility becomes nonzero. Figure~\ref{barriercases}(d) then shows $\nu(\omega,j_c)$ as a function of the probe frequency $\omega$ for different cavity coupling fractions $j_c/L =0.14, 0.71, 1$. When the cavity has not yet fully covered both ends, the visibility is zero across all frequencies $\omega$. Visibility only begins to distinguish between parities when the cavity fraction $j_c/L \approx 1$, as in the case without a barrier. 

With the barrier isolating the QD, the previous zero‐energy Andreev mode is lifted to finite energies ($\epsilon_0/\Delta_{\rm min} = 0.52$). This suggests that, based on the location of the absorption peaks, the two parities are already distinct in terms of energy (see Fig.~\ref{Appendix_Barrier_ImChi_W} in Appendix \ref{ImChi_W_AllCases}). Consequently, the disparity in amplitude and the related visibility are no longer significant; hence, we see that the visibility is nonzero well before it fully couples to the wire, and the energy itself distinguishes the parities. This is due to the fact that visibility is only clearly defined when both parities pertain to the same transition energies. These reasons lead to the visibility plots shown in Fig.~\ref{barriercases}(b) and Fig.~\ref{barriercases}(e), where it is evident that the parities can be effectively differentiated before the cavity fully couples with the entire wire. 

For the setup supporting QMBSs, the energy of the lowest mode remains pinned to zero energy even in the presence of a barrier ($\epsilon_0/\Delta_{\rm min} = 9.87\times 10^{-6}$), with the corresponding states localized at the two edges of the local topological segment within the SC. Figure~\ref{barriercases}(c) illustrates the visibility with respect to $j_c/L$, highlighting that it effectively differentiates the parity when the cavity is coupled to the QMBSs peaks of $|\psi_{L,R}(j)|^2$ at the boundaries of the topological region. 
In  Fig.~\ref{barriercases}(f), we plot $\nu(\omega,j_c)$ as a function of probe frequency $\omega$ for different coupling lengths of the cavity  $j_c/L=0.14, 0.71, 1$. Because the QMBS remain localized at the local topological edges (near $j_c/L = 0.14$
and $j_{\zeta}/L = 0.64$) of the topological region, visibility becomes nonzero already near $j_c/L \approx 0.71$, but well before it fully covers the SC part.

The results thus far demonstrate that microwave absorption visibility offers a clear and robust signature of Majorana nonlocality. Across all scenarios considered—pristine nanowires, wires coupled to QDs, or systems with disorder and tunnel barriers—the defining criterion remains unchanged: Finite visibility arises only when the cavity simultaneously couples to both Majorana end states. Crucially, this does not require coupling to the entire wire. In practice, it is sufficient for the cavity field to overlap with the wire ends, where the zero-energy bound states are localized. This allows for experimental implementations in which the cavity couples selectively to one or both QD regions at the ends of the wire, rather than to the entire proximitized segment. In particular, if an additional QD is present at the right end, then simultaneous coupling to both QDs provides direct access to the nonlocal correlations between the associated bound states. This unifying perspective sets the stage for the next section, where we apply the same criterion to the analytically tractable case of ``poor man’s'' Majoranas.

\section{Visibility of Poor Man's Majoranas}
\label{poormanMajorana}

In a minimal model, PMMs appear as zero-energy states in a two-site Kitaev chain Hamiltonian~\cite{Kitaev_Phys.-Usp..2001Unpaired} at the sweet spot when the tunneling is tuned to the superconducting pairing~\cite{Leijnse_Phys.Rev.B.2012Parity}. Multiple theoretical works have further addressed the QD-based platform for implementing PMMs
~\cite{Fulga_NewJ.Phys..2013Adaptive, Liu_Phys.Rev.Lett..2022Tunable,
Tsintzis_Phys.Rev.B.2022Creating,
Miles_Phys.Rev.B.2024Kitaev,
Samuelson_Phys.Rev.B.2024Minimal,
Tsintzis_PRXQuantum.2024Majorana,
Souto_Phys.Rev.Research.2023Probing,
Liu_CommunPhys.2024Enhancing,
TorresLuna_SciPostPhys.Core.2024Fluxtunable,
Liu_Phys.Rev.B.2024Coupling,
Pino_Phys.Rev.B.2024Minimal,
SeoaneSouto_NewTrendsandPlatformsforQuantumTechnologies.2024Subgap,
Gomez-Leon_Phys.Rev.B.2025Majorana,
Luethi_Phys.Rev.B.2024Perfect,
Luethi_Phys.Rev.B.2025Fate} with Majorana polarization proposed to quantify their quality
~\cite{Tsintzis_Phys.Rev.B.2022Creating,SeoaneSouto_NewTrendsandPlatformsforQuantumTechnologies.2024Subgap}. 

Experimentally, there has also been remarkable progress in implementing PMMs based on QDs coupled via a SC. The original proposal has been implemented in an  InSb nanowire hosting two QDs defined by electrostatic gates 
\cite{Dvir_Nature.2023Realization}. The differential conductance from each end showed simultaneous zero-bias peaks that split in phase as the sweet spot was detuned, matching PMM theory; nonlocal correlations between the peaks ruled out simple ABS explanations.  Reference
~\cite{Bordin_Nat.Nanotechnol..2025Enhanced}
extended the chain to three QDs, demonstrating that a finite mini-gap appears and, crucially, that the zero-bias peaks persist over a much wider gate-voltage window than in the dimer. A complementary approach replaced ordinary QD orbitals with Yu–Shiba–Rusinov  levels formed in a proximitised InAs/Al hybrid 
~\cite{Zatelli_NatCommun.2024Robust}, demonstrating markedly lower gate-noise sensitivity—another route to practical PMMs. Very recently, the implementation of PMM's was extended to two dimensions 
\cite{TenHaaf_Nature.2024Twosite}. Most of these works are rooted in quantum transport to probe the nature of the zero modes. Here we show that subjecting the PMMs to microwaves, using the visibility framework described in the previous sections, allows for complementary ways to characterize their properties and, moreover, to extract analytical expressions. Very recently, a similar approach has been implemented in Ref.~\cite{vanLoo_.2025Singleshot}.

\begin{figure}[t]
    \centering
    \includegraphics[clip, trim=4cm 2cm 4cm 2cm, width=\linewidth]{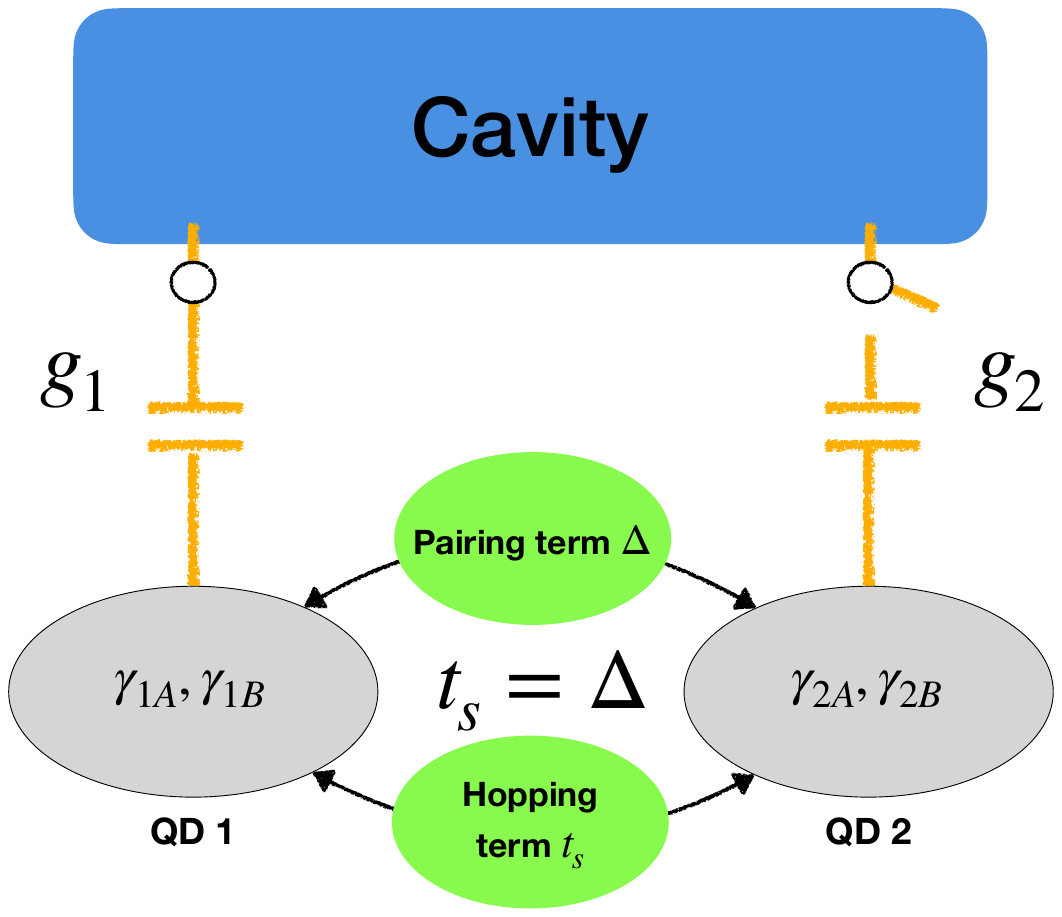}
    \caption{Two QDs (depicted in gray) are connected to a common SC (not shown). When the cross-Andreev reflection is balanced with the elastic cotunneling processes (i.e., when $t_s=\Delta$), a pair of Majorana fermions, $\gamma_{1A,2B}$, appear on the left and right dots, respectively. These fermions form a nonlocal zero-energy mode, referred to as the PMMs. The remaining two Majorana modes, $\gamma_{1B,2A}$, form a fermion with finite energy.  Each QD is  coupled to a microwave cavity via capacitors that can be turned on and off with $g_{1,2}$ as the cavity-dot coupling strength.}
    \label{poormanmajoranafigure}
\end{figure}

A sketch that describes the QD platform PMMs in the presence of the cavity is depicted in Fig.~\ref{poormanmajoranafigure}. The minimal model Hamiltonian that describes this scheme~\cite{Leijnse_Phys.Rev.B.2012Parity} can be written as $ H = H_{d} + H_{c} + H_{d-c}$, with
\begin{align}
 H_{d}&= \tilde{\epsilon}_1 n_1 + \tilde{\epsilon}_2 n_2 + (t_s d_1^\dagger d_2 + \Delta d_1^\dagger d_2^\dagger + {\rm H.c.})\nonumber\,,  \\
  H_{d-c}&= (g_1 n_1 + g_2 n_2)(a^\dagger + a)\nonumber\nonumber\,,
  \label{General2QDcoupledviaSCHamiltonian}
\end{align}
where $\tilde{\epsilon}_{i=1,2}$ is the on-site energy, $ n_i = d_i^\dagger d_i $ is the occupation number operator for QD \( i=1,2 \) with $d_i^\dagger$ ($d_i$) as the electron creation (annihilation) operator,   \( t_s \) is the tunneling amplitude, and \( \Delta \) is the superconducting pairing potential induced by the cross-Andreev reflection between the two QDs. The latter  are assumed to be capacitively coupled to a cavity with frequency $\omega_c$ and coupling strengths $g_{i=1,2}$. We stress that in order for the crossed Andreev reflection to be relevant, the length of the SC linking them needs to be less than the SC coherence length.  Defining $\Psi = \begin{pmatrix} d_1 & d_2 & d_1^\dagger & d_2^\dagger \end{pmatrix}^T$ enables us to cast $H_d$ in the form of $ H_d = \frac{1}{2}  \Psi^\dagger H_\text{BdG} \Psi$, where
\begin{align}
    H_{\rm BdG}=\left(
    \begin{array}{cccc}
\tilde{\epsilon}_1 & t_s& 0 & \Delta\\
t_s & \tilde{\epsilon}_2& -\Delta & 0\\
0 & -\Delta & -\tilde{\epsilon}_1 & -t_s\\
\Delta & 0 & -t_s & -\tilde{\epsilon}_2
    \end{array}
   \right) 
\end{align}
represent the BdG Hamiltonian. For $\tilde{\epsilon}_1 = \tilde{\epsilon}_2 = \tilde{\epsilon}$, the spectrum is given by
\begin{equation}
    E_{\pm}^{(\pm)}=\pm\sqrt{\tilde{\epsilon}^2+(t_s\pm\Delta)^2}\,,
\end{equation}
with two of the eigenvalues vanishing simultaneously when $\tilde{\epsilon}=0$ and $t_s=\Delta$. At that sweet spot, the QD pair hosts two zero-energy MBSs, exactly like the ends of a Kitaev chain with two sites.

We can define the self-adjoint Majorana operators $\gamma_{1A}=(d_{1}+d_{1}^\dagger)/{\sqrt{2}}$, $\gamma_{2B}=(d_{2}-d_{2}^\dagger)/{(i\sqrt{2})}$, and their partners $\tilde{\gamma}_{1B}=(d_{1}-d_{1}^\dagger)/{(i\sqrt{2})}$ and $\tilde{\gamma}_{2A}=(d_{2}+d_{2}^\dagger)/{\sqrt{2}}$ such that the many-body Hamiltonian at the sweet spot is  $H_d=-it_s\tilde{\gamma}_{1B}\tilde{\gamma}_{2A}$, while  $[\gamma_{1A(2B)},H_d]=0$. Hence, $\gamma_{1A}$ and $\gamma_{2B}$ drop out of the Hamiltonian and are located on the left and right QD, respectively. Consequently, the twofold degenerate  subspace is spanned by the states  
$\ket{G_e} = \left(  \ket{00} -\ket{11}     \right)/\sqrt{2}$ and 
$\ket{G_o} =  \left(  \ket{10} -\ket{01}     \right)/\sqrt{2}$, pertaining to the even ($e$) and odd ($o$) parity, respectively. The excited states, which are separated from the low-energy sector by $2\Delta$, are given by 
$\ket{E_e} =  \left(  \ket{00}+\ket{11}     \right)/\sqrt{2}$ and 
$\ket{E_o}=  \left(  \ket{10} +\ket{01}     \right)/\sqrt{2}$. In this eigenbasis, the QD-photon interaction becomes
\begin{align}
       H_{d-c}&=\frac{1}{2}\left[ - g_+
\ket{G_e}\bra{E_e}
%
+ g_-\,
\ket{G_o}\bra{E_o}
+{\rm h.c.}
\right](a^\dagger+a)\,,
\end{align}
with $g_{\pm}=g_1\pm g_2$. By tracing back through the derivations presented in prior sections, one arrives at the expression for electronic susceptibility as follows:
\begin{align}
    \chi_{e,o}(\omega)= & -\frac{1}{2}\frac{g_{\mp}^2 t_s}{4t_s^2 - \omega^2}\,.
    \label{qd-ph}
\end{align}
We see that $\chi_e(\omega)=0$ when $g_1=g_2$, while $\chi_e(\omega)=\chi_o(\omega)$ if either $g_1=0$ or $g_2=0$ (i.e., the cavity couples locally to either of the dots). These results are in agreement with  the recent findings in Ref.~\cite{vanLoo_.2025Singleshot} which focus on the measurement of the quantum capacitance to discriminate the two parities. Similar to previous discussions, the susceptibility can be divided into real and imaginary components (consider $\omega\rightarrow\omega+i\eta$). This time, let us concentrate on the real component of the susceptibility, which is relevant for non-demolition measurements when $\omega\ll\Delta$. Drawing a parallel to earlier sections, a visibility, denoted as $\tilde{\nu}$, can be defined in relation to the real component as follows:
\begin{equation}
    \tilde{\nu}\equiv\frac{\chi'_o(\omega)-\chi'_e(\omega)}{\chi'_o(\omega)+\chi'_e(\omega)}=\frac{2g_1 g_2}{g_1^2 + g_2^2}\,.
\end{equation}
The expression indicates that visibility becomes nonzero solely when the cavity interacts with both QDs, underscoring the nonlocal aspect of the PMMs.

\subsection*{Dynamical preparation of a specific parity state}

In this section, we briefly discuss a possible extension of the nonlocal coupling to the cavity: Preparation of a given parity state. In all previous sections, we considered the cavity's impact on the electronic system within the linear response framework. However, we now expand the discussion to scenarios where a strong cavity drive significantly alters the Majorana populations. For simplicity, we only focus here on PMMs, building on a double-$\Lambda$ four-level configuration  proposed for cooling atoms to their ground state \cite{Vacanti_NewJ.Phys..2009Cooling} (see also 
Ref.~\cite{Wesdorp_Phys.Rev.Lett..2023Dynamical}). However, the arguments can be generalized for more complicated setups and Hamiltonians, which we leave for future work. The scheme works as follows: Two near-degenerate ground states $|1\rangle$ and $|2\rangle$ (the low-energy manifold) are coupled
via coherent drivings to two excited states $|3\rangle$ and $|4\rangle$, respectively \cite{Vacanti_NewJ.Phys..2009Cooling}. Spontaneous or engineered dissipation provides cross-relaxation: state $|4\rangle$ decays into $|1\rangle$, and $|3\rangle$ decays into $|2\rangle$. This forms a closed cycle: 
\begin{itemize}
\item $|1\rangle$ is driven up to $|3\rangle$, which falls to $|2\rangle$
\item $|2\rangle$ is driven up to $|4\rangle$, which falls to $|1\rangle$\,.
\end{itemize}
By applying an {\it asymmetric} driving field to the two transitions, i.e., with different coupling strengths, one can induce a unidirectional population flow into the ground-state manifold, ultimately populating a selected target state. For example, suppose we drive the $|1\rangle\to|3\rangle$ transition resonantly (or with a slight detuning), while the $|2\rangle\to|4\rangle$ drive is turned off. Any population initially in $|1\rangle$ gets excited to $|3\rangle$ and then decays to $|2\rangle$, pumping $|1\rangle$ into $|2\rangle$. Likewise, any population in $|4\rangle$ will decay to $|1\rangle$, and then that will be pumped to $|2\rangle$ by the drive. In this regime, $|2\rangle$ acts as a dark reservoir state: It is easy to reach but hard to leave since we are not driving $|2\rangle$ out. The net effect is that all population is collected in $|2\rangle$. By choosing which laser to apply, one can pump towards either ground state. 

In order to implement (theoretically) this strategy within the PMMs framework, let us assume that the cavity is driven coherently at frequency $\Omega$, such that  $a\rightarrow \xi_d e^{-i\Omega t}$ and $a^\dagger\rightarrow\xi_d^* e^{i\Omega t}$, where  $\xi_d$  represents the complex amplitude of the cavity field set by the applied driving field. Referring to the previous section, we can designate the eigenstates as follows: $|1\rangle\equiv|G_e\rangle$, $|2\rangle\equiv|G_o\rangle$, $|3\rangle\equiv|E_o\rangle$, and $|4\rangle\equiv|E_e\rangle$. Next, we assume that once  the drive excites a quasiparticle in an excited energy state, it decays (escapes) directly into a fermion reservoir at a rate $\Gamma$ 
\cite{San-Jose_Phys.Rev.X.2015Majorana}. Furthermore, we assume that the ground state is equilibrated at a rate $\gamma\ll\Gamma$, induced, for example, by quasiparticle poisoning. The time-dependent density matrix $\rho(t)$ evolves as governed by:
\begin{align}
    \dot{\rho}(t) = [\mathcal{L}_{\rm coh}(t) + \mathcal{L}_{\rm diss}] \rho(t)\,,
\end{align}
where $\mathcal{L}_{\rm coh}$ and $\mathcal{L}_{\rm diss}$ represent the coherent and dissipative Liouvilleans, respectively. These are defined, respectively, as follows:
\begin{align}
    \mathcal{L}_{\rm coh}(t)\odot &= -i[H_{d} + H_{d-c}(t), \odot]\,,\nonumber\\
    \mathcal{L}_{\rm diss}\odot &= \sum_{j=1}^3\gamma_i\left(L_i\odot L_i^\dagger - \frac{1}{2}\{L_i^\dagger L_i, \odot\}\right)\,,
\end{align}
where $L_{1,2,3} = \tilde{d}_1, \tilde{d}_1^\dagger, \tilde{d}_2$ are the respective jump operators and $\tilde{d}_1 = \frac{1}{2}(d_1 + d_2 +  d_1^\dagger  -d_2^\dagger), \quad 
\tilde{d}_2 = \frac{1}{2}(d_1 + d_2 - d_1^\dagger +d_2^\dagger)$, where $i=1$ corresponds to the zero-energy state. Note that $\tilde{d}_2^\dagger$ does not appear, as its corresponding rate is taken to be zero in our scheme. The damping rates are given by $\gamma_1 =\gamma_2=\gamma$ and $\gamma_3  = \Gamma$, respectively. Using the unitary transformation $U(t) =\exp[-i\frac{\Omega t}{2}\sum_{p=e,o}\, \left(\ket{E_p}\bra{E_p} -\ket{G_p}\bra{G_p}\right) ]$ to move to a frame rotating at frequency $\Omega$, we get
\begin{align}
    \tilde{H}_{d-c}(t)=&
-\frac{g_+}{2}\Big[
\xi_d e^{2i\Omega t}\ket{E_e}\bra{G_e}
+\xi_d\ket{G_e}\bra{E_e}\Big] \nonumber\\
&\quad +\frac{g_-}{2}\Big[
\xi_d e^{2i\Omega t}\ket{E_o}\bra{G_o}
+\xi_d\ket{G_o}\bra{E_o}
\Big] +{\rm H.c.}
\end{align}
Using the rotating-wave approximation, i.e., discarding the fast-oscillating terms, we obtain
\begin{align}
\tilde{H}_{d-c}^{\rm RWA}=&
- \frac{g_+}{2} \, \xi_d \ket{G_e}\bra{E_e}
+ \frac{g_-}{2} \, \xi_d \ket{G_o}\bra{E_o} + {\rm H.c.},
\end{align}
while  $t_s\rightarrow \tilde{t}_s=t_s-\frac{\Omega}{2}$ 
in the double-QD Hamiltonian $H_d$.  The stationary density matrix, $\tilde{\rho}_s$, is determined from the condition $[\tilde{\mathcal{L}}_{\rm coh}(t)+\tilde{\mathcal{L}}_{\rm diss}]\tilde{\rho}_s=0$  in the new frame and can be solved exactly for this simple model. The population of the two lowest states of opposite parity (assuming $\xi_d$ to be real), 
$p_{G_e(G_o)}\equiv\langle G_e(G_o)|\tilde{\rho}_s|G_e(G_o)\rangle$ becomes
\begin{align}
p_{G_e(G_o)}&=\frac{\tilde{g}_{-(+)}^2\left(\Gamma^2+\tilde{g}_{+(-)}^2+16\tilde{t}_h^2\right)}{(16\tilde{t}_s^2+\Gamma^2)\left(\tilde{g}_-^2+\tilde{g}_+^2\right)+(2\tilde{g}_-\tilde{g}_+)^2}\,,
\end{align}
while the excited state populations 
$p_{E_o(E_e)}\equiv\langle E_o(E_e)|\tilde{\rho}_s|E_o(E_e)\rangle$ read: 
\begin{align}
p_{E_o}=p_{E_e}&=\frac{(\tilde{g}_-\tilde{g}_+)^2}{(16\tilde{t}_s^2+\Gamma^2)\left(\tilde{g}_-^2+\tilde{g}_+^2\right)+(2\tilde{g}_-\tilde{g}_+)^2}\,.
\end{align}
Above, we assumed $\gamma\sim0$ for simplicity and defined $\tilde{g}_{-(+)} = \xi_d g_{-(+)}$ and  $\tilde{g}_i= \xi_d g_i $. 
We see that 
 $p_{G_e}=p_{G_o}$ when one of the QDs is not coupled to the cavity (say, $\tilde{g}_2=0$), while 
 $p_{G_e}=p_{G_o}=1/2$ when the system is not driven, as expected for an incoherently populated degenerate ground state. However, for any $\tilde{g}_1\neq \tilde{g}_2$, the population of one of the lowest parity states becomes the largest, while in the limit $\tilde{g}_1=+(-) \tilde{g}_2$, we find 
$p_{G_o(G_e)}=1$, and the populations of all the other states vanish. Hence, when the cavity couples to both PMMs, it is possible to dynamically initialize the state with a given parity in the absence of any splitting within the lowest-energy subspace. Such a scheme (or variants of it) could be implemented for longer chains with  PMMs, or in mesoscopic spin-orbit SC Rashba nanowires, as long as the energy separation between the excited bulk levels is larger than their linewidths.

\section{Conclusions and outlook}
\label{conclusion}

In this work, we introduced the use of microwave absorption visibility in a one-dimensional  Rashba nanowire as a nonlocal method to differentiate MBSs from trivial zero-energy states, including ABSs localized on a nearby-QD  and QMBSs originating from uncontrolled inhomogeneous potentials. Through examining how microwave absorption visibility is affected by the cavity–wire coupling, we showed that true MBSs exhibit finite visibility only if both Majorana end modes are coupled to the cavity at the same time, highlighting their inherent nonlocal nature. In contrast, ABSs and QMBSs produce finite visibility without requiring full coverage, thereby providing a clear diagnostic criterion. We further established that this distinction persists in the presence of potential barriers and disorder, highlighting the robustness of the proposed probe. Finally, we showed that the same visibility framework applies to PMMs in QD–SC systems, where analytical results can be obtained and experimentally tested.

Looking ahead, several promising directions emerge. First, moving beyond the weak-coupling regime considered here, it would be intriguing to explore the strong-coupling limit, where hybrid excitations in the form of Majorana–polariton modes may arise. Second, interfacing the visibility framework with transport measurements in out-of-equilibrium settings could provide complementary insights into parity dynamics and decoherence, bridging microwave and dc probes. Third, extending the analysis to include additional dissipation channels, in particular quasiparticle poisoning, is essential for assessing the fidelity of cavity-based parity readout in realistic devices. Fourth, our approach naturally points to active cavity control schemes: By exploiting parity-selective visibility, the cavity could be used not only for nondemolition readout but also for implementing quantum gates with both nanowire MBSs and PMMs, paving the way toward cavity-assisted topological quantum information processing 
\cite{Contamin_npjQuantumInf.2021Hybrid}. Finally, an exciting avenue is to investigate situations with multiple MBSs, either along a single nanowire or in networks of coupled wires. In such setups, visibility could provide a versatile diagnostic for detecting nonlocal correlations among several MBSs, and possibly even visualize their dynamics.

Together, these directions highlight the versatility of cavity microwave absorption visibility as both a diagnostic and a control tool in Majorana platforms, offering a powerful complement to transport-based probes and opening new opportunities for the manipulation of topological bound states in hybrid superconducting systems.

\section{ACKNOWLEDGMENTS}

We acknowledge useful discussions with Peixin Shen, Daniel Loss, and Jelena Klinovaja. This work is supported by the ERC grant (Q-Light-Topo, Grant Agreement No. 101116525) (O. D.), by the Foundation for Polish Science project MagTop (No. FENG.02.01IP.05-0028/23), co-financed by the European Union from the funds of Priority 2 of the European Funds for a Smart Economy Program 2021–2027 (FENG) (S. P.  and M. T.), by the National Science Centre (Poland) OPUS Grant No. 2021/41/B/ST3/04475 (M. T.), and by the NAWA Bekker Grant No. BPN/BEK/2024/1/00310 (Poland) (M. T.).

\clearpage
\onecolumngrid
\appendix
\section{Evaluating the charge susceptibility in the Keldysh framework}
\label{App_Keldysh}
Here we provide a detailed description of the Keldysh approach used to compute susceptibility in the most general conditions that deviate from equilibrium. The contour-ordered response function reads 
\cite{Haug_.2008Quantum}:
\begin{align}
    \chi_C(\tau,\tau')&=-i\langle T_C[O_c(\tau)O_c(\tau')]\rangle
    =-i\sum_{p\sigma,j\sigma'}g_pg_j\langle T_C[c_{p\sigma}^\dagger(\tau) c_{p\sigma}(\tau)c_{j\sigma'}^\dagger(\tau') c_{j\sigma'}(\tau'))]\rangle\nonumber\\
    &=-i\sum_{p\sigma,j\sigma'}g_pg_j[G^C_{p\sigma,j\sigma'}(\tau,\tau')G^C_{j\sigma',p\sigma}(\tau',\tau)+F^C_{p\sigma,j\sigma'}(\tau,\tau')\bar{F}^{C}_{j\sigma',p\sigma}(\tau',\tau)]\,,
\end{align}
where $G_{p\sigma,j\sigma'}^C(\tau,\tau')=-i\langle T_C[c_{p\sigma}(\tau)c_{j\sigma'}^\dagger(\tau')]\rangle$ is the contour-ordered normal Green's function (GF), while $F_{j\sigma',p\sigma}^C(\tau,\tau')=-i\langle T_C[c_{j\sigma'}^\dagger(\tau)c_{p\sigma}^\dagger(\tau')]\rangle$ represents the anomalous one (with $\bar{F}\equiv F^*$). Using the Langreth rules 
\cite{Haug_.2008Quantum}, we can readily determine the retarded susceptibility in Eq.~\ref{susc} as follows:
\begin{align}
    \chi(t,t')&=-i\sum_{p\sigma,j\sigma'}g_jg_p[G_{p\sigma,j\sigma'}^<(t,t')G_{j\sigma',p\sigma}^a(t',t)+G_{p\sigma,j\sigma'}^r(t,t')G_{j\sigma',p\sigma}^<(t',t)\nonumber\\
&-F_{p\sigma,j\sigma'}^<(t,t')\bar{F}_{j\sigma',p\sigma}^a(t',t)-F_{p\sigma,j\sigma'}^r(t,t')\bar{F}_{j\sigma',p\sigma}^<(t',t)]\,.
\end{align}
Next, we assume the system is time-translation invariant, in which case all correlators depend only on time differences. Hence:
\begin{align}
    \chi(\Omega)&=-i\sum_{p\sigma,j\sigma'}g_jg_p\int_{-\infty}^\infty d\omega [G_{p\sigma,j\sigma'}^<(\omega)G_{j\sigma',p\sigma}^a(\omega+\Omega)+G_{p\sigma,j\sigma'}^r(\omega)G_{j\sigma',p\sigma}^<(\omega+\Omega)\nonumber\\
&-F_{p\sigma,j\sigma'}^<(\omega)\bar{F}_{j\sigma',p\sigma}^a(\omega+\Omega)-F_{p\sigma,j\sigma'}^r(\omega)\bar{F}_{j\sigma',p\sigma}^<(\omega+\Omega)]\,.
\end{align}
To make progress, let us explicitly write the retarded, advanced, and lesser GFs, respectively: 
\begin{align}
    G_{p\sigma,j\sigma'}^{r,a}(\omega)&=\sum_{n}\left(\frac{u^*_{n\sigma}(p)u_{n\sigma'}(j)}{\omega-\epsilon_n\pm i\eta}+\frac{v^*_{n\sigma}(p)v_{n\sigma'}(j)}{\omega+\epsilon_n\pm i\eta}\right)\,,\\
    G_{p\sigma,j\sigma'}^<(\omega)&=\sum_{n}\left[u^*_{n\sigma}(p)u_{n\sigma'}(j)f(\epsilon_n)\delta(\omega-\epsilon_n)+v^*_{n\sigma}(p)v_{n\sigma'}(j)(1-f(\epsilon_n))\delta(\omega+\epsilon_n)\right]\,,\\
    F_{p\sigma,j\sigma'}^{r,a}(\omega)&=\sum_{n}\left(\frac{u^*_{n\sigma}(p)v^*_{n\sigma'}(j)}{\omega-\epsilon_n\pm i\eta}+\frac{v^*_{n\sigma}(p)u^*_{n\sigma'}(j)}{\omega+\epsilon_n\pm i\eta}\right)\,,\\
    F_{p\sigma,j\sigma'}^<(\omega)&=\sum_{n}\left[u^*_{n\sigma}(p)v^*_{n\sigma'}(j)f(\epsilon_n)\delta(\omega-\epsilon_n)+v^*_{n\sigma}(p)u^{*}_{n\sigma'}(j)(1-f(\epsilon_n))\delta(\omega+\epsilon_n)\right]\,,
\end{align}
where $\eta>0$ is a small positive number that quantifies the lifetime of the levels. Thus, the real-time susceptibility in frequency space is
\begin{align}
    &    \chi(\Omega)
=\sum_{m\geq 0, n\geq 0}\Bigg\{  |\mathcal{M}_{nm}^{e}(j_c)|^2 \frac{ [f(\epsilon_m)-f(\epsilon_n)] }{\Omega+\epsilon_m-\epsilon_n +i \eta}  
+ \frac{|\mathcal{M}_{nm}^{o}(j_c)|^2}{2}\, \frac{[f(\epsilon_n)+f(\epsilon_m)-1]}{\Omega+\epsilon_n+\epsilon_m +i \eta}  \,    
+\frac{|\mathcal{M}_{nm}^{o}(j_c)|^2}{2}\, \frac{ [ 1-f(\epsilon_n)-f(\epsilon_m)] }{\Omega-\epsilon_n-\epsilon_m +i \eta}\,   \Bigg\},
\end{align}
where 
\begin{align}
\mathcal{M}^e_{nm}(j_c)&\!=\!\!\sum_{p=1}^{j_c}\sum_{\sigma=\uparrow,\downarrow}
\left[
u_{m\sigma}^*(p) u_{n\sigma}(p) -v_{m\sigma}^*(p) v_{n\sigma}(p)
\right]\nonumber\,,\\
\mathcal{M}^o_{nm}(j_c) &\!=\!\!\sum_{p=1}^{j_c}\sum_{\sigma=\uparrow,\downarrow} 
\left[
v_{m\sigma}(p) u_{n\sigma}(p)-u_{m\sigma}(p) v_{n\sigma}(p)
\right]\,.
\end{align}
Hence, we can establish several relationships between the lesser and the retarded/advanced GFs:
\begin{align}
    G_{p\sigma,j\sigma'}^<(\omega)&=[f(\omega)\theta(\omega)+(1-f(-\omega))\theta(-\omega)][G_{p\sigma,j\sigma'}^r(\omega)-G_{p\sigma,j\sigma'}^a(\omega)]\,,
\end{align}
and similarly for the anomalous components. We proceed as follows: (i) separate the summation so that terms involving the $n=0$ state are isolated, (ii) neglect bulk–bulk transitions, and (iii) apply the rotating-wave approximation by discarding all terms with resonances at $\omega<0$. All figures in the main text are generated using the full susceptibility expression given above. Putting everything together in the expression for the susceptibility and changing $\Omega \to \omega$, we obtain  Eq.~\ref{susc_general} in the main text (keeping only the terms that can lead to dissipation). 

\onecolumngrid
\section{Details of disorder}

\subsection{Chemical potential profile in the presence of disorder}
\label{MuProfileForDisorder}

\begin{figure*}[h]
    \centering

    \begin{minipage}[t]{0.32\linewidth}
        \centering
        \begin{tikzpicture}
            \node[anchor=south west, inner sep=0] (image) at (0,0) {
                \includegraphics[width=\columnwidth]{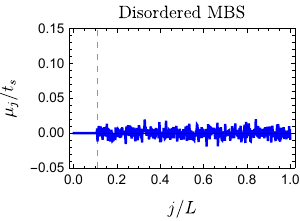}
            };
            \begin{scope}[x={(image.south east)}, y={(image.north west)}]
                \node[anchor=north west, font=\small\bfseries] at (0.03,1.02) {(a)};
            \end{scope}
        \end{tikzpicture}
    \end{minipage}
    \hfill
    \begin{minipage}[t]{0.32\linewidth}
        \centering
        \begin{tikzpicture}
            \node[anchor=south west, inner sep=0] (image) at (0,0) {
                \includegraphics[width=\linewidth]{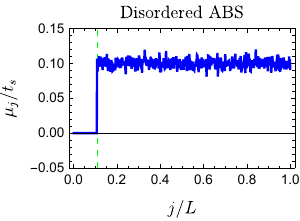}
            };
            \begin{scope}[x={(image.south east)}, y={(image.north west)}]
                \node[anchor=north west, font=\small\bfseries] at (0.03,1.02) {(b)};
            \end{scope}
        \end{tikzpicture}
    \end{minipage}
    \hfill
    \begin{minipage}[t]{0.32\linewidth}
        \centering
        \begin{tikzpicture}
            \node[anchor=south west, inner sep=0] (image) at (0,0) {
                \includegraphics[width=\linewidth]{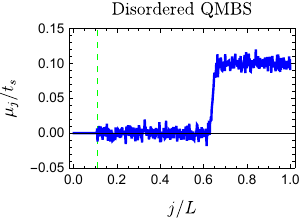}
            };
            \begin{scope}[x={(image.south east)}, y={(image.north west)}]
                \node[anchor=north west, font=\small\bfseries] at (0.03,1.02) {(c)};
            \end{scope}
        \end{tikzpicture}
    \end{minipage}

    \caption{Example of a disorder realization: the chemical potential profile is shown for Gaussian disorder with zero mean and standard deviation $\sigma_{\rm dis}=\Delta/2$, as described in the main text. The dashed green vertical line indicates the position of the QD–SC interface.}
    \label{Mu_DisorderPlots}
\end{figure*}

Here we illustrate the effect of Gaussian disorder on the spatial dependence of the chemical potential $\mu_j$. Representative realizations are shown in Fig.~\ref{Mu_DisorderPlots}, where random fluctuations around the clean profile distort the effective local band bottom. These plots are used in the main text to correlate disorder strength with the appearance of in-gap states and to evaluate the robustness of the visibility criterion. 

\subsection{Visibility for a different realization of disorder}
\label{Appendix_Disorder_2}

\begin{figure*}[t]
    \centering
    \begin{minipage}[t]{0.32\linewidth}
        \centering
        \begin{tikzpicture}
            \node[anchor=south west, inner sep=0] (image) at (0,0) {
                \includegraphics[width=\linewidth]{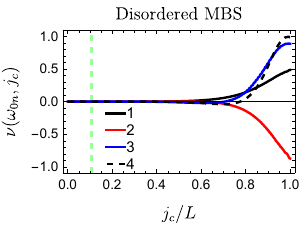}
            };
            \begin{scope}[x={(image.south east)}, y={(image.north west)}]
                \node[anchor=north west, font=\small\bfseries] at (0.03,1.02) {(a)};
            \end{scope}
        \end{tikzpicture}
    \end{minipage}
    \hfill
    \begin{minipage}[t]{0.32\linewidth}
        \centering
        \begin{tikzpicture}
            \node[anchor=south west, inner sep=0] (image) at (0,0) {
                \includegraphics[width=\linewidth]{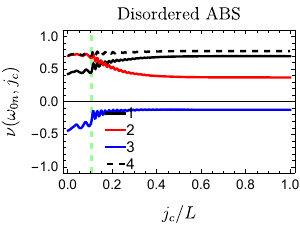}
            };
            \begin{scope}[x={(image.south east)}, y={(image.north west)}]
                \node[anchor=north west, font=\small\bfseries] at (0.03,1.02) {(b)};
            \end{scope}
        \end{tikzpicture}
    \end{minipage}
    \hfill
    \begin{minipage}[t]{0.32\linewidth}
        \centering
        \begin{tikzpicture}
            \node[anchor=south west, inner sep=0] (image) at (0,0) {
                \includegraphics[width=\linewidth]{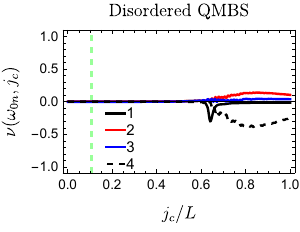}
            };
            \begin{scope}[x={(image.south east)}, y={(image.north west)}]
                \node[anchor=north west, font=\small\bfseries] at (0.03,1.02) {(c)};
            \end{scope}
        \end{tikzpicture}
    \end{minipage}  


    \begin{minipage}[t]{0.32\linewidth}
        \centering
        \begin{tikzpicture}
            \node[anchor=south west, inner sep=0] (image) at (0,0) {
                \includegraphics[width=\linewidth]{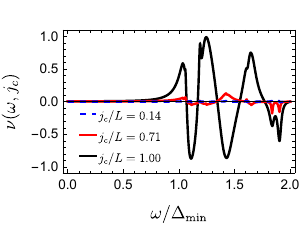}
            };
            \begin{scope}[x={(image.south east)}, y={(image.north west)}]
                \node[anchor=north west, font=\small\bfseries] at (0.03,1.02) {(d)};
            \end{scope}
        \end{tikzpicture}
    \end{minipage}
    \hfill
    \begin{minipage}[t]{0.32\linewidth}
        \centering
        \begin{tikzpicture}
            \node[anchor=south west, inner sep=0] (image) at (0,0) {
                \includegraphics[width=\linewidth]{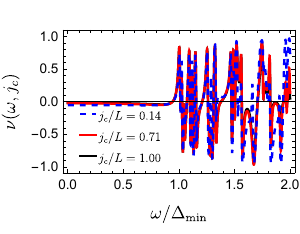}
            };
            \begin{scope}[x={(image.south east)}, y={(image.north west)}]
                \node[anchor=north west, font=\small\bfseries] at (0.03,1.02) {(e)};
            \end{scope}
        \end{tikzpicture}
    \end{minipage}
    \hfill
    \begin{minipage}[t]{0.32\linewidth}
        \centering
        \begin{tikzpicture}
            \node[anchor=south west, inner sep=0] (image) at (0,0) {
                \includegraphics[width=\linewidth]{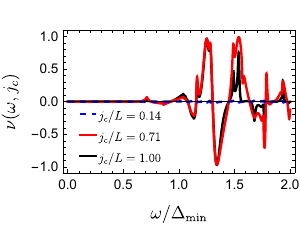}
            };
            \begin{scope}[x={(image.south east)}, y={(image.north west)}]
                \node[anchor=north west, font=\small\bfseries] at (0.03,1.02) {(f)};
            \end{scope}
        \end{tikzpicture}
    \end{minipage}

    \caption{A different Gaussian disorder realization: Microwave absorption properties of a topological SC nanowire with an adjacent QD for a different disorder ($\sigma_{\rm dis}=\Delta/2$) realization. [(a)--(c)] Visibility $\nu(\omega_{0n},j_{c})$ as a function of fraction of wire $j_c/L$ covered by cavity for MBSs, ABSs, and QMBSs, respectively, for the transitions to the excited states $n=1,2,3,4$. [(d)--(f)] Visibility $\nu(\omega,j_{c})$ as function of probe frequency $\omega$ for different fraction of wire coupled by cavity for MBSs, ABSs, and QMBSs, respectively. This shows that, unlike the ABS and QMBS, the visibility $\nu(\omega_{0n},j_{c})$  of MBS vanishes unless the cavity couples to both the Majorana modes at the SC edges, a signature for MBSs nonlocality. The parameters utilized are presented in Table \ref{parametertable} for each case.
    }
    \label{Disorder_2}
\end{figure*}

To confirm the generality of our conclusions in the main text, Fig.~\ref{Disorder_2} depicts the visibility results for alternative Gaussian disorder realizations. We compare MBSs, ABSs, and QMBSs under identical conditions/parameters. In all cases, the behavior discussed in Sec.\ref{EffectOfDisorder} remains unchanged: The MBSs visibility requires simultaneous coupling to both ends, while the ABSs and QMBSs produce finite visibility already for partial coverage. These complementary plots demonstrate that the qualitative distinctions are not sensitive to the specific disorder configuration.


\section{Scaling of Majorana coherence length and visibility with sites}\label{scaling_dicussion}
\begin{figure*}[h]
    \centering

    \begin{minipage}[t]{0.32\linewidth}
        \centering
        \begin{tikzpicture}
            \node[anchor=south west, inner sep=0] (image) at (0,0) {
                \includegraphics[width=\columnwidth]{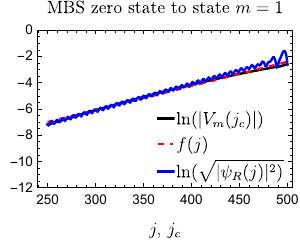}
            };
            \begin{scope}[x={(image.south east)}, y={(image.north west)}]
                \node[anchor=north west, font=\small\bfseries] at (0.0,1.02) {(a)};
            \end{scope}
        \end{tikzpicture}
    \end{minipage}
    \hfill
    \begin{minipage}[t]{0.32\linewidth}
        \centering
        \begin{tikzpicture}
            \node[anchor=south west, inner sep=0] (image) at (0,0) {
                \includegraphics[width=\linewidth]{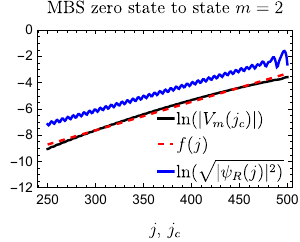}
            };
            \begin{scope}[x={(image.south east)}, y={(image.north west)}]
                \node[anchor=north west, font=\small\bfseries] at (0.0,1.02) {(b)};
            \end{scope}
        \end{tikzpicture}
    \end{minipage}
    \hfill
    \begin{minipage}[t]{0.32\linewidth}
        \centering
        \begin{tikzpicture}
            \node[anchor=south west, inner sep=0] (image) at (0,0) {
                \includegraphics[width=\linewidth]{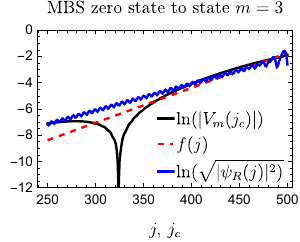}
            };
            \begin{scope}[x={(image.south east)}, y={(image.north west)}]
                \node[anchor=north west, font=\small\bfseries] at (0.0,1.02) {(c)};
            \end{scope}
        \end{tikzpicture}
    \end{minipage}

    \caption{In all panels, the blue curve represents $\ln\big(|\psi_R(j)|^2\,\big)$ of the Majorana zero-energy state, which has a Majorana coherence length $\xi_{\rm MBS} \approx 48$ and amplitude $A \approx 0.12$, as determined (fitted function for MBS is not shown in plots) from the fit function $f(j) = \frac{j}{\xi_{\rm MBS}} + \ln(A) - \frac{L_s}{\xi_{\rm MBS}}$. The black curve in each panel corresponds to $V_m(j_c)=\ln\big(|\,|\mathcal{M}^o_{m0}(j_c)|^2 - |\mathcal{M}^e_{m0}(j_c)|^2\,|\big)$ for the transitions indicated at the top of the panel. We consider $V_m(j_c)$ instead of the actual visibility $\nu(\omega_{0n}, j_c)$ as the former does not depend on linewidth $\eta$. As the significant part of the $|\psi_R(j)|$ is closer to the right edge, fitting of $V_m(j_c)$ was performed over the site range $[L_s/2, L_s]$, and the resulting fit function $f(j)$ for $V_m(j_c)$ is shown as the red dashed line.}
    \label{Vis_WFR_Fits}
\end{figure*}

In the main text, we quoted the Majorana coherence length in the pristine case as $\xi_{\rm MBS}\approx 48$ (in units of the lattice spacing $a=1$). Here we explicitly demonstrate that both the Majorana wavefunctions and the microwave visibility 
decay exponentially and that the corresponding decay lengths are essentially the same. The coherence length $\xi_{\rm MBS}$ is extracted numerically from the exponential tails of the Majorana wavefunctions along the nanowire, i.e., as a function of the position index $j$. By contrast, the visibility is analyzed as a function of the cavity coverage $j_c$, 
namely the last site coupled to the cavity. 
Figure~\ref{Vis_WFR_Fits} illustrates that the exponential decay in both cases is characterized by very similar length scales, providing direct evidence that the visibility faithfully tracks the spatial decay of the MBSs themselves. 

\onecolumngrid
\section{Plots of imaginary part of susceptibility for the cases discussed in main text}
\label{ImChi_W_AllCases}

\subsection{Imaginary part of the susceptibility}

\begin{figure*}[h]
    \centering
    \begin{minipage}[t]{0.32\linewidth}
        \centering
        \begin{tikzpicture}
            \node[anchor=south west, inner sep=0, outer sep=0] (image) at (0,0) {
                \includegraphics[width=\linewidth]{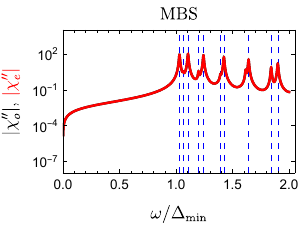}
            };
            \begin{scope}[x={(image.south east)}, y={(image.north west)}]
                \node[anchor=north west, font=\small\bfseries] at (0.03,1.02) {(a)};
            \end{scope}
        \end{tikzpicture}
    \end{minipage}
    \hfill
    \begin{minipage}[t]{0.32\linewidth}
        \centering
        \begin{tikzpicture}
            \node[anchor=south west, inner sep=0, outer sep=0] (image) at (0,0) {
                \includegraphics[width=\linewidth
                ]{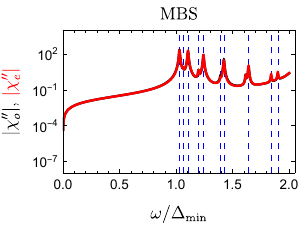}
            };
            \begin{scope}[x={(image.south east)}, y={(image.north west)}]
                \node[anchor=north west, font=\small\bfseries] at (0.03,1.02) {(b)};
            \end{scope}
        \end{tikzpicture}
    \end{minipage}
    \hfill
    \begin{minipage}[t]{0.32\linewidth}
        \centering
        \begin{tikzpicture}
            \node[anchor=south west, inner sep=0, outer sep=0] (image) at (0,0) {
                \includegraphics[width=\linewidth]{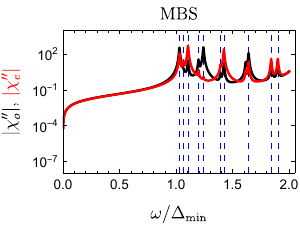}
            };
            \begin{scope}[x={(image.south east)}, y={(image.north west)}]
                \node[anchor=north west, font=\small\bfseries] at (0.03,1.02) {(c)};
            \end{scope}
        \end{tikzpicture}
    \end{minipage}
    
    
    \begin{minipage}[t]{0.32\linewidth}
        \centering
        \begin{tikzpicture}
            \node[anchor=south west, inner sep=0, outer sep=0] (image) at (0,0) {
                \includegraphics[width=\linewidth]{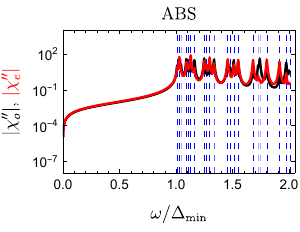}
            };
            \begin{scope}[x={(image.south east)}, y={(image.north west)}]
                \node[anchor=north west, font=\small\bfseries] at (0.03,1.02) {(d)};
            \end{scope}
        \end{tikzpicture}
    \end{minipage}
    \hfill
    \begin{minipage}[t]{0.32\linewidth}
        \centering
        \begin{tikzpicture}
            \node[anchor=south west, inner sep=0, outer sep=0] (image) at (0,0) {
                \includegraphics[width=\linewidth]{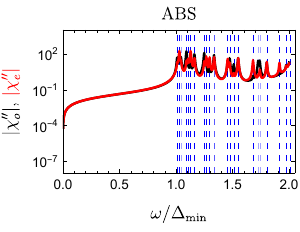}
            };
            \begin{scope}[x={(image.south east)}, y={(image.north west)}]
                \node[anchor=north west, font=\small\bfseries] at (0.03,1.02) {(e)};
            \end{scope}
        \end{tikzpicture}
    \end{minipage}
    \hfill
    \begin{minipage}[t]{0.32\linewidth}
        \centering
        \begin{tikzpicture}
            \node[anchor=south west, inner sep=0, outer sep=0] (image) at (0,0) {
                \includegraphics[width=\linewidth]{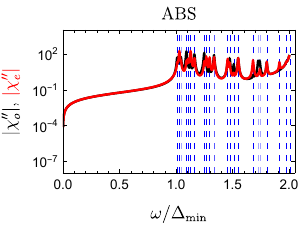}
            };
            \begin{scope}[x={(image.south east)}, y={(image.north west)}]
                \node[anchor=north west, font=\small\bfseries] at (0.03,1.02) {(f)};
            \end{scope}
        \end{tikzpicture}
    \end{minipage}

    \begin{minipage}[t]{0.32\linewidth}
        \centering
        \begin{tikzpicture}
            \node[anchor=south west, inner sep=0, outer sep=0] (image) at (0,0) {
                \includegraphics[width=\linewidth]{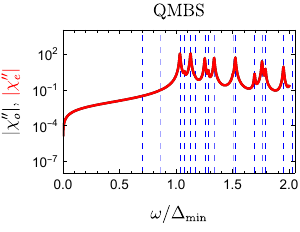}
            };
            \begin{scope}[x={(image.south east)}, y={(image.north west)}]
                \node[anchor=north west, font=\small\bfseries] at (0.03,1.02) {(g)};
            \end{scope}
        \end{tikzpicture}
    \end{minipage}
    \hfill
    \begin{minipage}[t]{0.32\linewidth}
        \centering
        \begin{tikzpicture}
            \node[anchor=south west, inner sep=0, outer sep=0] (image) at (0,0) {
                \includegraphics[width=\linewidth]{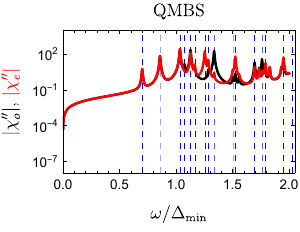}
            };
            \begin{scope}[x={(image.south east)}, y={(image.north west)}]
                \node[anchor=north west, font=\small\bfseries] at (0.03,1.02) {(h)};
            \end{scope}
        \end{tikzpicture}
    \end{minipage}
    \hfill
    \begin{minipage}[t]{0.32\linewidth}
        \centering
        \begin{tikzpicture}
            \node[anchor=south west, inner sep=0, outer sep=0] (image) at (0,0) {
                \includegraphics[width=\linewidth]{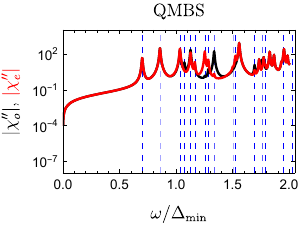}
            };
            \begin{scope}[x={(image.south east)}, y={(image.north west)}]
                \node[anchor=north west, font=\small\bfseries] at (0.03,1.02) {(i)};
            \end{scope}
        \end{tikzpicture}
    \end{minipage}

    \caption{
    The imaginary part of the susceptibility, as a function of the frequency $\omega$ for clean cases involving a QD, is detailed for MBSs [(a)--(c)], ABSs [(d)--(f)], and QMBSs [(g)--(i)]. Each column represents different cavity covering fractions $j_c/L$: Panels (a), (d), and (g) correspond to $j_{c}/L=0.14$; (b), (e), and (h) pertain to $j_{c}/L=0.71$; and (c), (f), and (i) relate to $j_{c}/L=1$. The blue dashed line indicates the resonant frequencies for transitions from the zero-energy state to the excited states.
    }
    \label{ImChi_W_AllCases_CleanQD}
\end{figure*}

\begin{figure*}[t]
    \centering
    \begin{minipage}[t]{0.32\linewidth}
        \centering
        \begin{tikzpicture}
            \node[anchor=south west, inner sep=0, outer sep=0] (image) at (0,0) {
                \includegraphics[width=\linewidth]{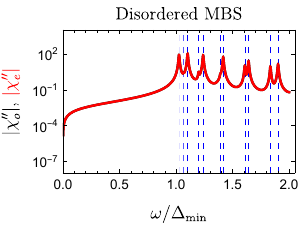}
            };
            \begin{scope}[x={(image.south east)}, y={(image.north west)}]
                \node[anchor=north west, font=\small\bfseries] at (0.03,1.02) {(a)};
            \end{scope}
        \end{tikzpicture}
    \end{minipage}
    \hfill
    \begin{minipage}[t]{0.32\linewidth}
        \centering
        \begin{tikzpicture}
            \node[anchor=south west, inner sep=0, outer sep=0] (image) at (0,0) {
                \includegraphics[width=\linewidth
                ]{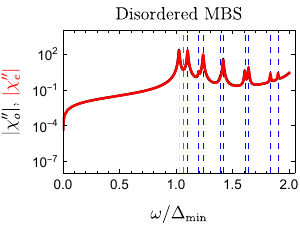}
            };
            \begin{scope}[x={(image.south east)}, y={(image.north west)}]
                \node[anchor=north west, font=\small\bfseries] at (0.03,1.02) {(b)};
            \end{scope}
        \end{tikzpicture}
    \end{minipage}
    \hfill
    \begin{minipage}[t]{0.32\linewidth}
        \centering
        \begin{tikzpicture}
            \node[anchor=south west, inner sep=0, outer sep=0] (image) at (0,0) {
                \includegraphics[width=\linewidth]{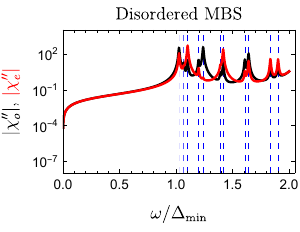}
            };
            \begin{scope}[x={(image.south east)}, y={(image.north west)}]
                \node[anchor=north west, font=\small\bfseries] at (0.03,1.02) {(c)};
            \end{scope}
        \end{tikzpicture}
    \end{minipage}
    
    
    \begin{minipage}[t]{0.32\linewidth}
        \centering
        \begin{tikzpicture}
            \node[anchor=south west, inner sep=0, outer sep=0] (image) at (0,0) {
                \includegraphics[width=\linewidth]{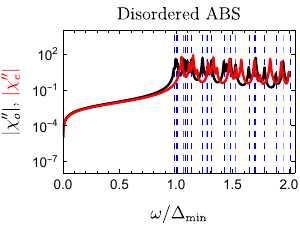}
            };
            \begin{scope}[x={(image.south east)}, y={(image.north west)}]
                \node[anchor=north west, font=\small\bfseries] at (0.03,1.02) {(d)};
            \end{scope}
        \end{tikzpicture}
    \end{minipage}
    \hfill
    \begin{minipage}[t]{0.32\linewidth}
        \centering
        \begin{tikzpicture}
            \node[anchor=south west, inner sep=0, outer sep=0] (image) at (0,0) {
                \includegraphics[width=\linewidth]{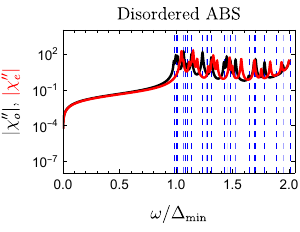}
            };
            \begin{scope}[x={(image.south east)}, y={(image.north west)}]
                \node[anchor=north west, font=\small\bfseries] at (0.03,1.02) {(e)};
            \end{scope}
        \end{tikzpicture}
    \end{minipage}
    \hfill
    \begin{minipage}[t]{0.32\linewidth}
        \centering
        \begin{tikzpicture}
            \node[anchor=south west, inner sep=0, outer sep=0] (image) at (0,0) {
                \includegraphics[width=\linewidth]{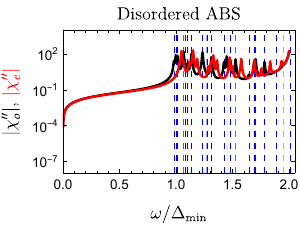}
            };
            \begin{scope}[x={(image.south east)}, y={(image.north west)}]
                \node[anchor=north west, font=\small\bfseries] at (0.03,1.02) {(f)};
            \end{scope}
        \end{tikzpicture}
    \end{minipage}

    \begin{minipage}[t]{0.32\linewidth}
        \centering
        \begin{tikzpicture}
            \node[anchor=south west, inner sep=0, outer sep=0] (image) at (0,0) {
                \includegraphics[width=\linewidth]{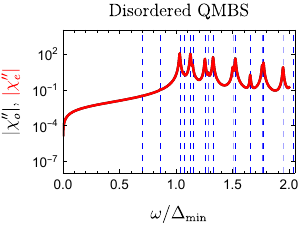}
            };
            \begin{scope}[x={(image.south east)}, y={(image.north west)}]
                \node[anchor=north west, font=\small\bfseries] at (0.03,1.02) {(g)};
            \end{scope}
        \end{tikzpicture}
    \end{minipage}
    \hfill
    \begin{minipage}[t]{0.32\linewidth}
        \centering
        \begin{tikzpicture}
            \node[anchor=south west, inner sep=0, outer sep=0] (image) at (0,0) {
                \includegraphics[width=\linewidth]{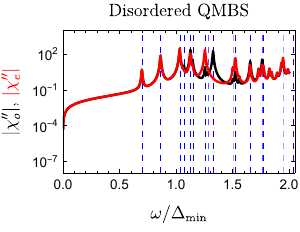}
            };
            \begin{scope}[x={(image.south east)}, y={(image.north west)}]
                \node[anchor=north west, font=\small\bfseries] at (0.03,1.02) {(h)};
            \end{scope}
        \end{tikzpicture}
    \end{minipage}
    \hfill
    \begin{minipage}[t]{0.32\linewidth}
        \centering
        \begin{tikzpicture}
            \node[anchor=south west, inner sep=0, outer sep=0] (image) at (0,0) {
                \includegraphics[width=\linewidth]{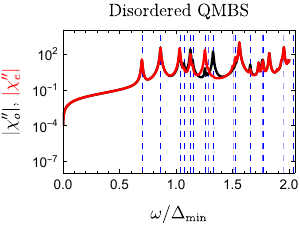}
            };
            \begin{scope}[x={(image.south east)}, y={(image.north west)}]
                \node[anchor=north west, font=\small\bfseries] at (0.03,1.02) {(i)};
            \end{scope}
        \end{tikzpicture}
    \end{minipage}

    \caption{
       The imaginary part of the susceptibility, as a function of the frequency $\omega$ for disordered cases involving a QD, is detailed for MBSs [(a)--(c)], ABSs [(d)--(f)], and QMBSs [(g)--(i)]. Each column represents different cavity covering fractions $j_c/L$: Panels (a), (d), and (g) correspond to $j_{c}/L=0.14$; (b), (e), and (h) pertain to $j_{c}/L=0.71$; and (c), (f), and (i) relate to $j_{c}/L=1$. The blue dashed line indicates the resonant frequencies for transitions from the zero-energy state to the excited states.
    }
    \label{ImChi_W_AllCases_Disordered}
\end{figure*}

\begin{figure*}[t]
    \centering
    \begin{minipage}[t]{0.32\linewidth}
        \centering
        \begin{tikzpicture}
            \node[anchor=south west, inner sep=0, outer sep=0] (image) at (0,0) {
                \includegraphics[width=\linewidth]{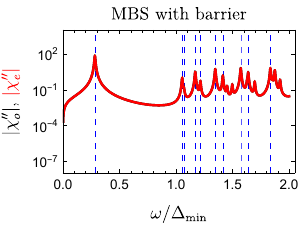}
            };
            \begin{scope}[x={(image.south east)}, y={(image.north west)}]
                \node[anchor=north west, font=\small\bfseries] at (0.03,1.02) {(a)};
            \end{scope}
        \end{tikzpicture}
    \end{minipage}
    \hfill
    \begin{minipage}[t]{0.32\linewidth}
        \centering
        \begin{tikzpicture}
            \node[anchor=south west, inner sep=0, outer sep=0] (image) at (0,0) {
                \includegraphics[width=\linewidth
                ]{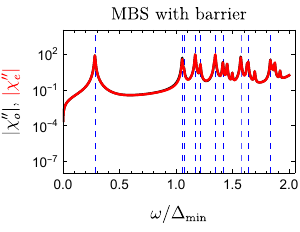}
            };
            \begin{scope}[x={(image.south east)}, y={(image.north west)}]
                \node[anchor=north west, font=\small\bfseries] at (0.03,1.02) {(b)};
            \end{scope}
        \end{tikzpicture}
    \end{minipage}
    \hfill
    \begin{minipage}[t]{0.32\linewidth}
        \centering
        \begin{tikzpicture}
            \node[anchor=south west, inner sep=0, outer sep=0] (image) at (0,0) {
                \includegraphics[width=\linewidth]{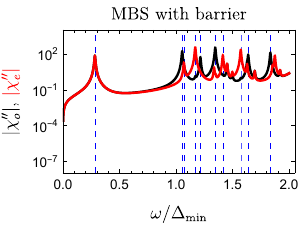}
            };
            \begin{scope}[x={(image.south east)}, y={(image.north west)}]
                \node[anchor=north west, font=\small\bfseries] at (0.03,1.02) {(c)};
            \end{scope}
        \end{tikzpicture}
    \end{minipage}
    
    
    \begin{minipage}[t]{0.32\linewidth}
        \centering
        \begin{tikzpicture}
            \node[anchor=south west, inner sep=0, outer sep=0] (image) at (0,0) {
                \includegraphics[width=\linewidth]{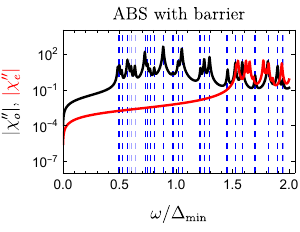}
            };
            \begin{scope}[x={(image.south east)}, y={(image.north west)}]
                \node[anchor=north west, font=\small\bfseries] at (0.03,1.02) {(d)};
            \end{scope}
        \end{tikzpicture}
    \end{minipage}
    \hfill
    \begin{minipage}[t]{0.32\linewidth}
        \centering
        \begin{tikzpicture}
            \node[anchor=south west, inner sep=0, outer sep=0] (image) at (0,0) {
                \includegraphics[width=\linewidth]{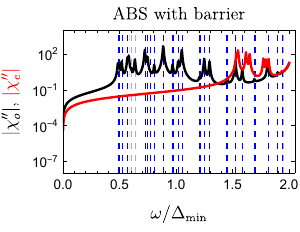}
            };
            \begin{scope}[x={(image.south east)}, y={(image.north west)}]
                \node[anchor=north west, font=\small\bfseries] at (0.03,1.02) {(e)};
            \end{scope}
        \end{tikzpicture}
    \end{minipage}
    \hfill
    \begin{minipage}[t]{0.32\linewidth}
        \centering
        \begin{tikzpicture}
            \node[anchor=south west, inner sep=0, outer sep=0] (image) at (0,0) {
                \includegraphics[width=\linewidth]{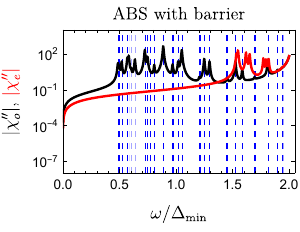}
            };
            \begin{scope}[x={(image.south east)}, y={(image.north west)}]
                \node[anchor=north west, font=\small\bfseries] at (0.03,1.02) {(f)};
            \end{scope}
        \end{tikzpicture}
    \end{minipage}

    \begin{minipage}[t]{0.32\linewidth}
        \centering
        \begin{tikzpicture}
            \node[anchor=south west, inner sep=0, outer sep=0] (image) at (0,0) {
                \includegraphics[width=\linewidth]{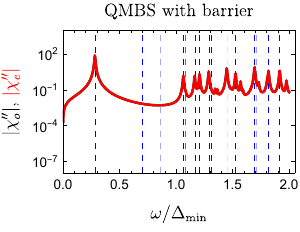}
            };
            \begin{scope}[x={(image.south east)}, y={(image.north west)}]
                \node[anchor=north west, font=\small\bfseries] at (0.03,1.02) {(g)};
            \end{scope}
        \end{tikzpicture}
    \end{minipage}
    \hfill
    \begin{minipage}[t]{0.32\linewidth}
        \centering
        \begin{tikzpicture}
            \node[anchor=south west, inner sep=0, outer sep=0] (image) at (0,0) {
                \includegraphics[width=\linewidth]{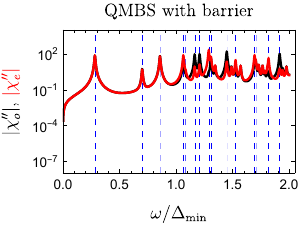}
            };
            \begin{scope}[x={(image.south east)}, y={(image.north west)}]
                \node[anchor=north west, font=\small\bfseries] at (0.03,1.02) {(h)};
            \end{scope}
        \end{tikzpicture}
    \end{minipage}
    \hfill
    \begin{minipage}[t]{0.32\linewidth}
        \centering
        \begin{tikzpicture}
            \node[anchor=south west, inner sep=0, outer sep=0] (image) at (0,0) {
                \includegraphics[width=\linewidth]{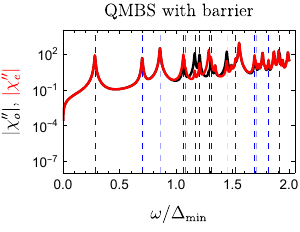}
            };
            \begin{scope}[x={(image.south east)}, y={(image.north west)}]
                \node[anchor=north west, font=\small\bfseries] at (0.03,1.02) {(i)};
            \end{scope}
        \end{tikzpicture}
    \end{minipage}
    \caption{With the barrier isolating the QD, the previous zero‐energy Andreev mode is lifted to finite energies ($\epsilon_0/\Delta_{\rm min} = 0.52$). The imaginary part of the susceptibility, as a function of the frequency $\omega$ for clean cases in the presence of a barrier, is detailed for MBSs [(a)--(c)], ABSs [(d)--(f)], and QMBSs [(g)--(i)]. Each column represents different cavity covering fractions $j_c/L$: Panels (a), (d), and (g) correspond to $j_{c}/L=0.14$; (b), (e), and (h) pertain to $j_{c}/L=0.71$; and (c), (f), and (i) relate to $j_{c}/L=1$. The blue dashed line indicates the resonant frequencies for transitions from the zero-energy state to the excited states.
    }
    \label{Appendix_Barrier_ImChi_W}
\end{figure*}

In this Appendix, we show representative results for the imaginary part of the electronic susceptibility in Fig.~\ref{ImChi_W_AllCases_CleanQD},
Fig.~\ref{ImChi_W_AllCases_Disordered}, and 
Fig.~\ref{Appendix_Barrier_ImChi_W}, 
which directly contribute to the renormalization of the cavity decay rate via $\kappa_{o,e}' = \kappa + 2\,\mathrm{Im}\,\chi_{o,e}(\omega_c)$, as discussed in Sec.~\ref{modelHamiltonian}. 
The imaginary component corresponds to absorption processes and is therefore closely linked to the visibility analysis presented in the main text. In the trivial case, the even- and odd-parity responses are nearly identical, leading to vanishing visibility. By contrast, in the topological regime with well-separated MBSs, the absorption spectra differ significantly between parities, giving rise to finite visibility. The resonance peaks in $\chi_{o,e}''(\omega)$ correspond to transitions from the zero-energy state into excited quasiparticle states, with their relative weight encoding the spatial overlap of Majorana components with the cavity-coupled region. 

These results reinforce the central message of the paper: 
the imaginary part of the susceptibility carries clear signatures of Majorana nonlocality and provides a direct microscopic origin for the parity-dependent visibility employed as a diagnostic tool in the main text.

\subsection{Real part of susceptibility}\label{ReChi_W_AllCases}
\begin{figure*}[h]
    \centering
    \begin{minipage}[t]{0.32\linewidth}
        \centering
        \begin{tikzpicture}
            \node[anchor=south west, inner sep=0, outer sep=0] (image) at (0,0) {
                \includegraphics[width=\linewidth]{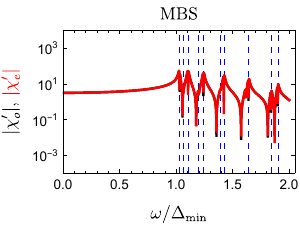}
            };
            \begin{scope}[x={(image.south east)}, y={(image.north west)}]
                \node[anchor=north west, font=\small\bfseries] at (0.03,1.02) {(a)};
            \end{scope}
        \end{tikzpicture}
    \end{minipage}
    \hfill
    \begin{minipage}[t]{0.32\linewidth}
        \centering
        \begin{tikzpicture}
            \node[anchor=south west, inner sep=0, outer sep=0] (image) at (0,0) {
                \includegraphics[width=\linewidth
                ]{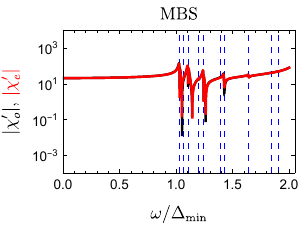}
            };
            \begin{scope}[x={(image.south east)}, y={(image.north west)}]
                \node[anchor=north west, font=\small\bfseries] at (0.03,1.02) {(b)};
            \end{scope}
        \end{tikzpicture}
    \end{minipage}
    \hfill
    \begin{minipage}[t]{0.32\linewidth}
        \centering
        \begin{tikzpicture}
            \node[anchor=south west, inner sep=0, outer sep=0] (image) at (0,0) {
                \includegraphics[width=\linewidth]{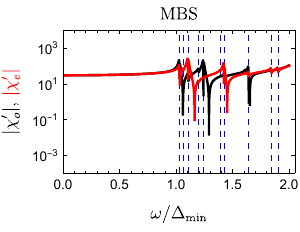}
            };
            \begin{scope}[x={(image.south east)}, y={(image.north west)}]
                \node[anchor=north west, font=\small\bfseries] at (0.03,1.02) {(c)};
            \end{scope}
        \end{tikzpicture}
    \end{minipage}
    
    
    \begin{minipage}[t]{0.32\linewidth}
        \centering
        \begin{tikzpicture}
            \node[anchor=south west, inner sep=0, outer sep=0] (image) at (0,0) {
                \includegraphics[width=\linewidth]{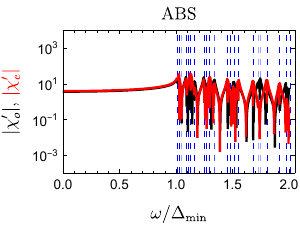}
            };
            \begin{scope}[x={(image.south east)}, y={(image.north west)}]
                \node[anchor=north west, font=\small\bfseries] at (0.03,1.02) {(d)};
            \end{scope}
        \end{tikzpicture}
    \end{minipage}
    \hfill
    \begin{minipage}[t]{0.32\linewidth}
        \centering
        \begin{tikzpicture}
            \node[anchor=south west, inner sep=0, outer sep=0] (image) at (0,0) {
                \includegraphics[width=\linewidth]{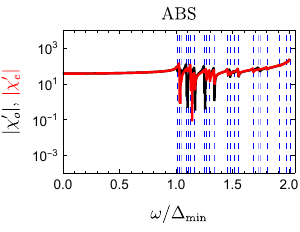}
            };
            \begin{scope}[x={(image.south east)}, y={(image.north west)}]
                \node[anchor=north west, font=\small\bfseries] at (0.03,1.02) {(e)};
            \end{scope}
        \end{tikzpicture}
    \end{minipage}
    \hfill
    \begin{minipage}[t]{0.32\linewidth}
        \centering
        \begin{tikzpicture}
            \node[anchor=south west, inner sep=0, outer sep=0] (image) at (0,0) {
                \includegraphics[width=\linewidth]{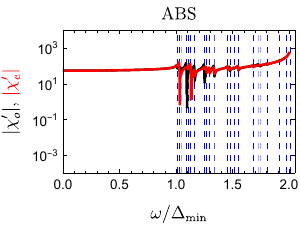}
            };
            \begin{scope}[x={(image.south east)}, y={(image.north west)}]
                \node[anchor=north west, font=\small\bfseries] at (0.03,1.02) {(f)};
            \end{scope}
        \end{tikzpicture}
    \end{minipage}

    \begin{minipage}[t]{0.32\linewidth}
        \centering
        \begin{tikzpicture}
            \node[anchor=south west, inner sep=0, outer sep=0] (image) at (0,0) {
                \includegraphics[width=\linewidth]{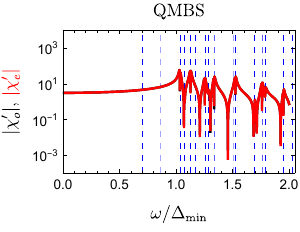}
            };
            \begin{scope}[x={(image.south east)}, y={(image.north west)}]
                \node[anchor=north west, font=\small\bfseries] at (0.03,1.02) {(g)};
            \end{scope}
        \end{tikzpicture}
    \end{minipage}
    \hfill
    \begin{minipage}[t]{0.32\linewidth}
        \centering
        \begin{tikzpicture}
            \node[anchor=south west, inner sep=0, outer sep=0] (image) at (0,0) {
                \includegraphics[width=\linewidth]{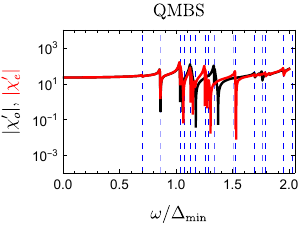}
            };
            \begin{scope}[x={(image.south east)}, y={(image.north west)}]
                \node[anchor=north west, font=\small\bfseries] at (0.03,1.02) {(h)};
            \end{scope}
        \end{tikzpicture}
    \end{minipage}
    \hfill
    \begin{minipage}[t]{0.32\linewidth}
        \centering
        \begin{tikzpicture}
            \node[anchor=south west, inner sep=0, outer sep=0] (image) at (0,0) {
                \includegraphics[width=\linewidth]{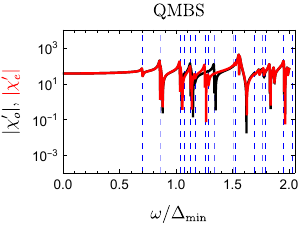}
            };
            \begin{scope}[x={(image.south east)}, y={(image.north west)}]
                \node[anchor=north west, font=\small\bfseries] at (0.03,1.02) {(i)};
            \end{scope}
        \end{tikzpicture}
    \end{minipage}

    \caption{
    Plots of the real part of susceptibility for different covering fractions for the clean cases with QD. Plots (a), (d), (g) show the plots for $j_{c}/L=0.14$, (b), (e), (h) are for  $j_{c}/L=0.71$ and (c), (f), (i) are for $j_{c}/L=1$. The blue dashed lines mark the resonant frequencies of transitions from zero energy state to excited states.
    }
    \label{ReChi_W_AllCases_CleanQD}
\end{figure*}

\begin{figure*}[t]
    \centering
    \begin{minipage}[t]{0.32\linewidth}
        \centering
        \begin{tikzpicture}
            \node[anchor=south west, inner sep=0, outer sep=0] (image) at (0,0) {
                \includegraphics[width=\linewidth]{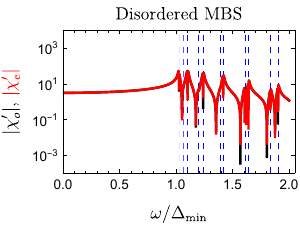}
            };
            \begin{scope}[x={(image.south east)}, y={(image.north west)}]
                \node[anchor=north west, font=\small\bfseries] at (0.03,1.02) {(a)};
            \end{scope}
        \end{tikzpicture}
    \end{minipage}
    \hfill
    \begin{minipage}[t]{0.32\linewidth}
        \centering
        \begin{tikzpicture}
            \node[anchor=south west, inner sep=0, outer sep=0] (image) at (0,0) {
                \includegraphics[width=\linewidth
                ]{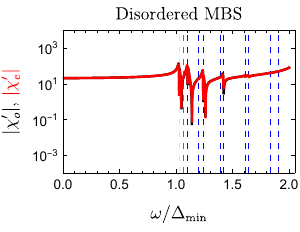}
            };
            \begin{scope}[x={(image.south east)}, y={(image.north west)}]
                \node[anchor=north west, font=\small\bfseries] at (0.03,1.02) {(b)};
            \end{scope}
        \end{tikzpicture}
    \end{minipage}
    \hfill
    \begin{minipage}[t]{0.32\linewidth}
        \centering
        \begin{tikzpicture}
            \node[anchor=south west, inner sep=0, outer sep=0] (image) at (0,0) {
                \includegraphics[width=\linewidth]{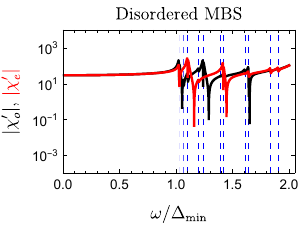}
            };
            \begin{scope}[x={(image.south east)}, y={(image.north west)}]
                \node[anchor=north west, font=\small\bfseries] at (0.03,1.02) {(c)};
            \end{scope}
        \end{tikzpicture}
    \end{minipage}
    
    
    \begin{minipage}[t]{0.32\linewidth}
        \centering
        \begin{tikzpicture}
            \node[anchor=south west, inner sep=0, outer sep=0] (image) at (0,0) {
                \includegraphics[width=\linewidth]{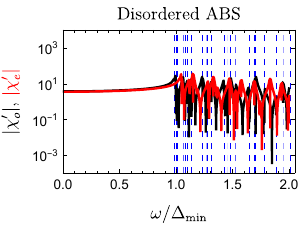}
            };
            \begin{scope}[x={(image.south east)}, y={(image.north west)}]
                \node[anchor=north west, font=\small\bfseries] at (0.03,1.02) {(d)};
            \end{scope}
        \end{tikzpicture}
    \end{minipage}
    \hfill
    \begin{minipage}[t]{0.32\linewidth}
        \centering
        \begin{tikzpicture}
            \node[anchor=south west, inner sep=0, outer sep=0] (image) at (0,0) {
                \includegraphics[width=\linewidth]{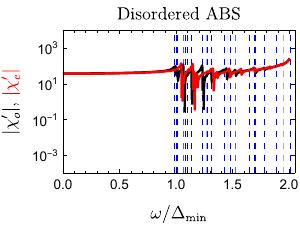}
            };
            \begin{scope}[x={(image.south east)}, y={(image.north west)}]
                \node[anchor=north west, font=\small\bfseries] at (0.03,1.02) {(e)};
            \end{scope}
        \end{tikzpicture}
    \end{minipage}
    \hfill
    \begin{minipage}[t]{0.32\linewidth}
        \centering
        \begin{tikzpicture}
            \node[anchor=south west, inner sep=0, outer sep=0] (image) at (0,0) {
                \includegraphics[width=\linewidth]{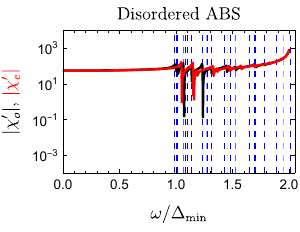}
            };
            \begin{scope}[x={(image.south east)}, y={(image.north west)}]
                \node[anchor=north west, font=\small\bfseries] at (0.03,1.02) {(f)};
            \end{scope}
        \end{tikzpicture}
    \end{minipage}

    \begin{minipage}[t]{0.32\linewidth}
        \centering
        \begin{tikzpicture}
            \node[anchor=south west, inner sep=0, outer sep=0] (image) at (0,0) {
                \includegraphics[width=\linewidth]{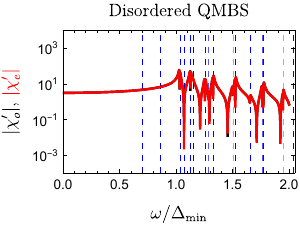}
            };
            \begin{scope}[x={(image.south east)}, y={(image.north west)}]
                \node[anchor=north west, font=\small\bfseries] at (0.03,1.02) {(g)};
            \end{scope}
        \end{tikzpicture}
    \end{minipage}
    \hfill
    \begin{minipage}[t]{0.32\linewidth}
        \centering
        \begin{tikzpicture}
            \node[anchor=south west, inner sep=0, outer sep=0] (image) at (0,0) {
                \includegraphics[width=\linewidth]{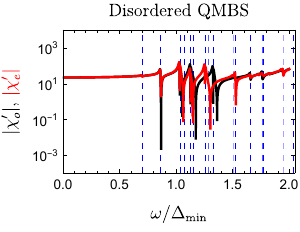}
            };
            \begin{scope}[x={(image.south east)}, y={(image.north west)}]
                \node[anchor=north west, font=\small\bfseries] at (0.03,1.02) {(h)};
            \end{scope}
        \end{tikzpicture}
    \end{minipage}
    \hfill
    \begin{minipage}[t]{0.32\linewidth}
        \centering
        \begin{tikzpicture}
            \node[anchor=south west, inner sep=0, outer sep=0] (image) at (0,0) {
                \includegraphics[width=\linewidth]{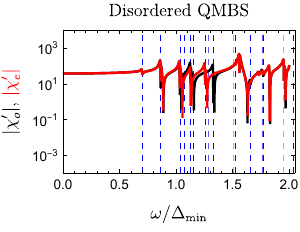}
            };
            \begin{scope}[x={(image.south east)}, y={(image.north west)}]
                \node[anchor=north west, font=\small\bfseries] at (0.03,1.02) {(i)};
            \end{scope}
        \end{tikzpicture}
    \end{minipage}

    \caption{
    Plot of real part of susceptibility for different covering fractions in the presence of disorder. Plots (a), (d), and (g) are for $j_{c}/L=0.14$; (b), (e), and (h) are for  $j_{c}/L=0.71$; and (c), (f), and (i) are for $j_{c}/L=1$. The blue dashed line marks the resonant frequencies of transitions from zero-energy state to excited states.
    }
    \label{ReChi_W_AllCases_Disordered}
\end{figure*}

\begin{figure*}[t]
    \centering
    \begin{minipage}[t]{0.32\linewidth}
        \centering
        \begin{tikzpicture}
            \node[anchor=south west, inner sep=0, outer sep=0] (image) at (0,0) {
                \includegraphics[width=\linewidth]{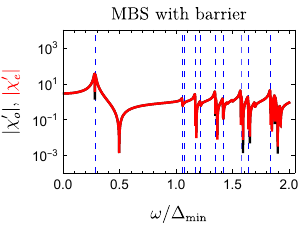}
            };
            \begin{scope}[x={(image.south east)}, y={(image.north west)}]
                \node[anchor=north west, font=\small\bfseries] at (0.03,1.02) {(a)};
            \end{scope}
        \end{tikzpicture}
    \end{minipage}
    \hfill
    \begin{minipage}[t]{0.32\linewidth}
        \centering
        \begin{tikzpicture}
            \node[anchor=south west, inner sep=0, outer sep=0] (image) at (0,0) {
                \includegraphics[width=\linewidth
                ]{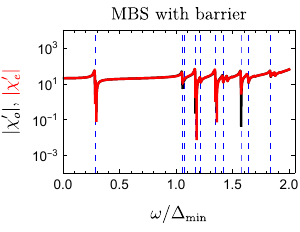}
            };
            \begin{scope}[x={(image.south east)}, y={(image.north west)}]
                \node[anchor=north west, font=\small\bfseries] at (0.03,1.02) {(b)};
            \end{scope}
        \end{tikzpicture}
    \end{minipage}
    \hfill
    \begin{minipage}[t]{0.32\linewidth}
        \centering
        \begin{tikzpicture}
            \node[anchor=south west, inner sep=0, outer sep=0] (image) at (0,0) {
                \includegraphics[width=\linewidth]{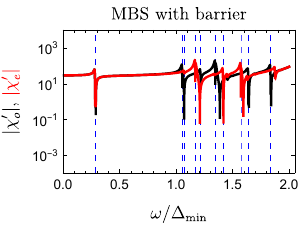}
            };
            \begin{scope}[x={(image.south east)}, y={(image.north west)}]
                \node[anchor=north west, font=\small\bfseries] at (0.03,1.02) {(c)};
            \end{scope}
        \end{tikzpicture}
    \end{minipage}
    
    
    \begin{minipage}[t]{0.32\linewidth}
        \centering
        \begin{tikzpicture}
            \node[anchor=south west, inner sep=0, outer sep=0] (image) at (0,0) {
                \includegraphics[width=\linewidth]{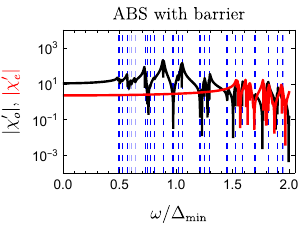}
            };
            \begin{scope}[x={(image.south east)}, y={(image.north west)}]
                \node[anchor=north west, font=\small\bfseries] at (0.03,1.02) {(d)};
            \end{scope}
        \end{tikzpicture}
    \end{minipage}
    \hfill
    \begin{minipage}[t]{0.32\linewidth}
        \centering
        \begin{tikzpicture}
            \node[anchor=south west, inner sep=0, outer sep=0] (image) at (0,0) {
                \includegraphics[width=\linewidth]{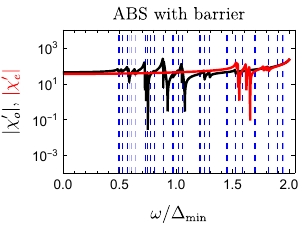}
            };
            \begin{scope}[x={(image.south east)}, y={(image.north west)}]
                \node[anchor=north west, font=\small\bfseries] at (0.03,1.02) {(e)};
            \end{scope}
        \end{tikzpicture}
    \end{minipage}
    \hfill
    \begin{minipage}[t]{0.32\linewidth}
        \centering
        \begin{tikzpicture}
            \node[anchor=south west, inner sep=0, outer sep=0] (image) at (0,0) {
                \includegraphics[width=\linewidth]{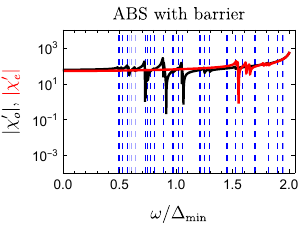}
            };
            \begin{scope}[x={(image.south east)}, y={(image.north west)}]
                \node[anchor=north west, font=\small\bfseries] at (0.03,1.02) {(f)};
            \end{scope}
        \end{tikzpicture}
    \end{minipage}

    \begin{minipage}[t]{0.32\linewidth}
        \centering
        \begin{tikzpicture}
            \node[anchor=south west, inner sep=0, outer sep=0] (image) at (0,0) {
                \includegraphics[width=\linewidth]{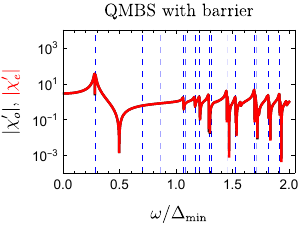}
            };
            \begin{scope}[x={(image.south east)}, y={(image.north west)}]
                \node[anchor=north west, font=\small\bfseries] at (0.03,1.02) {(g)};
            \end{scope}
        \end{tikzpicture}
    \end{minipage}
    \hfill
    \begin{minipage}[t]{0.32\linewidth}
        \centering
        \begin{tikzpicture}
            \node[anchor=south west, inner sep=0, outer sep=0] (image) at (0,0) {
                \includegraphics[width=\linewidth]{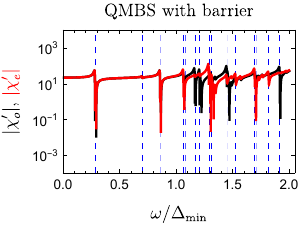}
            };
            \begin{scope}[x={(image.south east)}, y={(image.north west)}]
                \node[anchor=north west, font=\small\bfseries] at (0.03,1.02) {(h)};
            \end{scope}
        \end{tikzpicture}
    \end{minipage}
    \hfill
    \begin{minipage}[t]{0.32\linewidth}
        \centering
        \begin{tikzpicture}
            \node[anchor=south west, inner sep=0, outer sep=0] (image) at (0,0) {
                \includegraphics[width=\linewidth]{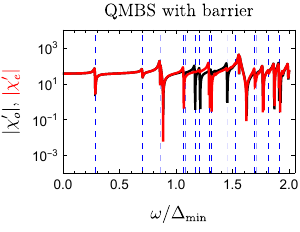}
            };
            \begin{scope}[x={(image.south east)}, y={(image.north west)}]
                \node[anchor=north west, font=\small\bfseries] at (0.03,1.02) {(i)};
            \end{scope}
        \end{tikzpicture}
    \end{minipage}

    \caption{Plots of the real part of susceptibility for different covering fractions in the presence of a barrier. Plots (a), (d), and (g) are for $j_{c}/L=0.14$; (b), (e), and (h) are for  $j_{c}/L=0.71$; and (c), (f), and (i) are for $j_{c}/L=1$. The blue dashed lines mark the resonant frequencies of transitions from zero-energy state to excited states.
    }
    \label{Appendix_Barrier_ReChi_W}
\end{figure*}

In this Appendix, we present the real part of the susceptibility, $\chi'_{e(o)}(\omega)=\mathrm{Re}\,\chi(\omega)$ in Fig.~\ref{ReChi_W_AllCases_CleanQD},
Fig.~\ref{ReChi_W_AllCases_Disordered}, and
Fig.~\ref{Appendix_Barrier_ReChi_W}, for the QD-coupled cases discussed in the main text. As noted there, $\chi'_{e(o)}$ renormalizes the cavity frequency $\omega_c$, shifting $\omega_c \to \omega_c+\chi'_{e(o)}(\omega_c)$. Similarly to the dissipative response, the MBSs parity-dependent dispersive shift occurs when the cavity couples to both MBSs. In contrast, for ABSs and QMBSs, a discernible shift emerges even for partial coupling, i.e., before the cavity fully couples to the entire wire.

\clearpage

\onecolumngrid
\section{Alternative case for quasi-Majorana}\label{NewQMBSAppendix}
\begin{figure*}[h]
    \centering

    \begin{minipage}[c]{0.32\linewidth}
        \centering
        \begin{tikzpicture}
            \node[anchor=north west, inner sep=0, outer sep=0] (image) at (0,0) {
                   \includegraphics[width=\linewidth,
    trim={4.5cm 1cm 5cm 2cm}, clip
    ]{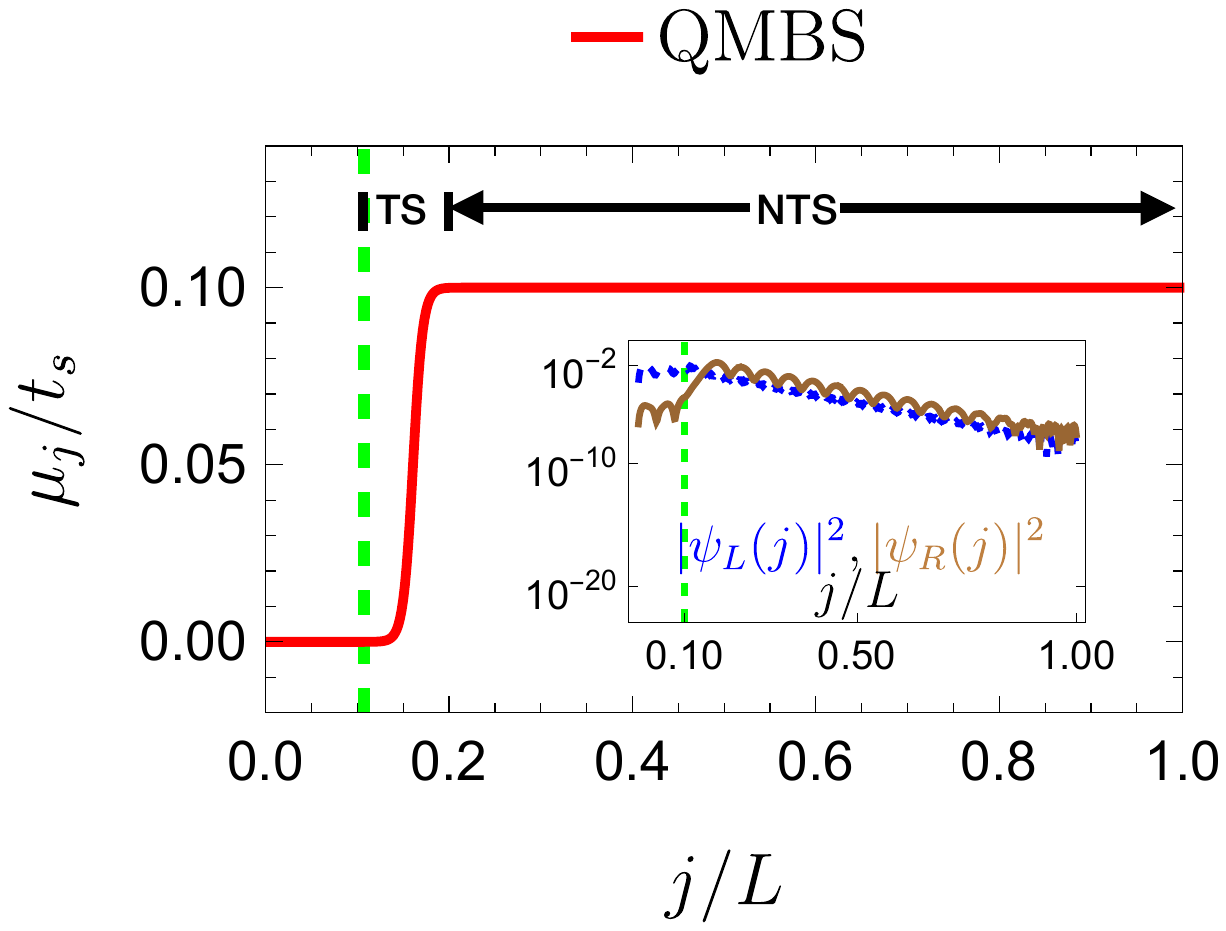}
            };
            \begin{scope}[x={(image.south east)}, y={(image.north west)}]
                \node[anchor=north west, font=\small\bfseries] at (0.03,1.02) {(a)};
            \end{scope}
        \end{tikzpicture}
    \end{minipage}
     \hfill
    \begin{minipage}[c]{0.32\linewidth}
        \centering
        \begin{tikzpicture}
            \node[anchor=south west, inner sep=0, outer sep=0] (image) at (0,0) {
                \includegraphics[width=\linewidth,
                 trim={0cm 0cm 0cm 0.4cm},clip]{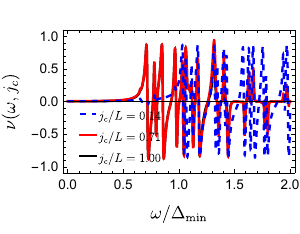}
            };
            \begin{scope}[x={(image.south east)}, y={(image.north west)}]
                \node[anchor=north west, font=\small\bfseries] at (0.03,1.1) {(b)};
            \end{scope}
        \end{tikzpicture}
    \end{minipage}
    \hfill
    \begin{minipage}[c]{0.32\linewidth}
        \centering
        \begin{tikzpicture}
            \node[anchor=south west, inner sep=0, outer sep=0] (image) at (0,0) {
                \includegraphics[ width=\linewidth,
    trim={0cm 0cm 0cm 0.4cm}, clip]{{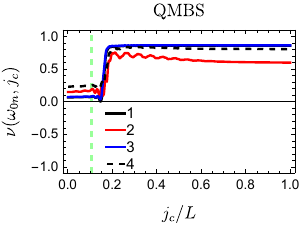}}
            };
            \begin{scope}[x={(image.south east)}, y={(image.north west)}]
                \node[anchor=north west, font=\small\bfseries] at (0.03,1.1) {(c)};
            \end{scope}
        \end{tikzpicture}
    \end{minipage}
    \caption{   
    An alternative scenario for QMBSs: Panel (a) shows the  chemical potential profile [Eq.~\eqref{ChemcialPotentialForArxivQMBSProfile}] marking the relative size of the topological superconducting (TS) part and the nontopological superconducting (NTS) part. The TS changes to NTS near $j_\xi/L = 0.16$ (other parameters same as QMBS in Table~\ref{parametertable}) and $\epsilon_0/\Delta_{\min} = 7.61 \times 10^{-3}$ . Due to the overlapping of the two QMBS peaks, the lowest energy is higher than the well-separated QMBS case in Fig.~\ref{QMBSMuProfile}. Panels (b) and (c) show the visibility as function of probe frequency $\omega$ and covering fraction $j_c/L$ respectively. Both (b) and (c) show that the visibility is well distinguished just as the cavity couples past  $j_\xi/L = 0.16$ as the cavity already couples to both peaks which are near $j_\xi/L = 0.16$. 
    }
       \label{NewQMBSAllResults}
\end{figure*}

As discussed in Sec.~\ref{VisibilityofzeroenergyQMBSs} of the main text, the precise form of QMBSs depends on the smoothness of the potential profile. In Fig.~\ref{NewQMBSAllResults}, we present an alternative configuration in which the two QMBS peaks move closer together as the system approaches the topological–trivial transition near $j_\xi/L = 0.16$, yielding a topological regime of $0.10 \leq j/L \leq 0.16$.

\twocolumngrid


\bibliographystyle{apsrev4-2}
 \bibliography{References}

\end{document}